\documentclass[conference,compsoc]{IEEEtran}

\usepackage{marvosym}
\usepackage{xspace}
\usepackage{amsmath}
\usepackage{pifont}
\usepackage{url}
\usepackage{listings}
\usepackage{hyperref}
\usepackage{cleveref}
\usepackage{booktabs}
\usepackage{multirow}
\usepackage{tabularx}
\usepackage{adjustbox}
\usepackage{tcolorbox}
\usepackage{xcolor}
\usepackage{graphicx}
\usepackage{comment}
\usepackage[linesnumbered,ruled,vlined]{algorithm2e}

\definecolor{codeGreen}{RGB}{0,100,0}
\definecolor{codeRed}{RGB}{163,21,21}
\definecolor{codeGray}{RGB}{110,110,110}
\lstdefinestyle{phpStyle}{
    language=PHP,
    basicstyle=\ttfamily\footnotesize,
    keywordstyle=\color{codeGreen}\bfseries,
    stringstyle=\color{codeRed},
    commentstyle=\color{codeGray},
    numbers=left,
    numberstyle=\ttfamily\scriptsize\color{gray},
    stepnumber=1,
    numbersep=3pt,
    frame=lines,
    rulecolor=\color{black},
    xleftmargin=1.4em,
    framexleftmargin=1.4em,
    backgroundcolor=\color{white},
    showstringspaces=false,
    captionpos=b,
    tabsize=4,
    breaklines=true,
    columns=fullflexible,
    keepspaces=true,
    morekeywords={class, function, public, protected, return, new, extends}
}
\crefname{lstlisting}{Listing}{Listings}
\Crefname{lstlisting}{Listing}{Listings}

\usepackage[labelformat=parens]{subcaption} 

\usepackage{enumitem}
\setlist{
    leftmargin=*
}

\newtcolorbox{rqbox}{
    sharp corners,
    boxrule=0pt,
    leftrule=2pt,
    colframe=black,
    colback=gray!10,
    left=3pt,
    right=3pt,
    top=3pt,
    bottom=3pt,
    before skip=8pt,
    after skip=8pt
}

\newcommand{\tool}{\textsc{ChainAgent}\xspace}

\ifCLASSOPTIONcompsoc
  \usepackage[nocompress]{cite}
\else
  \usepackage{cite}
\fi

\ifCLASSINFOpdf
\else
\fi

\begin{document}

\title{\tool: Automated Chainable Vulnerability Discovery in PHP Web Applications}


\author{
\IEEEauthorblockN{
Xinyi Wang\textsuperscript{1,2},
Yeting Li\textsuperscript{1,2,\Letter},
Shizhe Cui\textsuperscript{3},
Jiaming Guo\textsuperscript{1,2}
Qin Mai\textsuperscript{1,2},
Yecheng Sun\textsuperscript{1,2},\\
Hengyu Yang\textsuperscript{1,2},
Yujie Sun\textsuperscript{1,2}
Huina Chao\textsuperscript{1,2},
Yang Xiao\textsuperscript{1,2},
Feng Li\textsuperscript{1,2},
Wei Huo\textsuperscript{1,2}
}
\IEEEauthorblockA{
\textsuperscript{1}\{CAS-KLONAT, BKLONSPT\}, Institute of Information Engineering, CAS
\quad
\textsuperscript{2}School of Cyber Security, UCAS\\
\textsuperscript{3}China Southern Power Grid Digital Grid Group Information and Telecommunication Technology Co., Ltd.
}
}

\maketitle

\begingroup
\renewcommand{\thefootnote}{\Letter}
\footnotetext{Yeting Li is the corresponding author.}
\endgroup

\begin{abstract}
High-impact attacks against PHP web applications often emerge from chaining vulnerabilities that provide limited capabilities with other vulnerabilities or intended program behaviors. Discovering such chains requires identifying not only individual flaws, but also the dependencies that allow their capabilities to be composed. Existing vulnerability discovery techniques primarily analyze flaws in isolation, while automated attack-planning approaches typically assume that the relevant attack components are already known. This separation makes it difficult to discover previously unknown components whose security relevance becomes apparent only through their potential roles in an exploit chain.

We present \tool, a goal-directed LLM-agent framework that jointly discovers exploit-chain components and constructs their capability dependencies from application source code. Guided by an empirical study of 386 real-world web exploit chains, \tool represents vulnerabilities and functional gadgets--intended program behaviors that can be repurposed in attacks--through preconditions and capabilities over security-relevant objects. Rather than discovering vulnerabilities first and constructing attack paths afterward, \tool uses an evolving exploit graph to guide source-code inspection. It searches backward from unmet preconditions for potential capability producers and forward from discovered capabilities for functional gadgets that consume or amplify them. Object-centered chaining knowledge derived from our empirical study further directs these searches toward promising vulnerability classes, functional operations, and object accesses.

We evaluate \tool against Codex, a general-purpose coding agent, on 15 PHP applications across three shared LLM backends. \tool achieves 42.02\%--44.63\% recall for goal-reaching components and recovers up to 4.4 times as many correct goal-reaching dependency edges as Codex. An additional evaluation of 30 real-world PHP applications identifies 20 vulnerabilities and 13 exploit chains, nine of which achieve remote code execution (RCE).

\end{abstract}


%
\IEEEpeerreviewmaketitle

\section{Introduction}
\label{sec:introduction}

PHP remains a widely deployed server-side programming language, powering over 69.9\% of websites with known server-side technologies~\cite{w3techs_php_2026}. Their widespread deployment, Internet-facing exposure, and frequent processing of untrusted inputs make PHP web applications an attractive attack surface. Vulnerabilities in these applications can expose sensitive information, compromise access controls, and, in severe cases, enable remote code execution (RCE) or complete server takeover. Identifying and mitigating such vulnerabilities before exploitation remains a critical challenge in web application security.

Extensive research has investigated automated vulnerability discovery in PHP web applications~\cite{DBLP:conf/uss/ZIPPER, luo2022tchecker, DBLP:conf/uss/XSSky, DBLP:journals/pacmpl/ARTEMIS, DBLP:conf/sp/WangCXWGWDLW24, DBLP:journals/tse/PHPJoy, icse/CoBrA, DBLP:conf/sp/LiMZWL24, DBLP:conf/sp/Witcher, DBLP:conf/ndss/FUSE, DBLP:conf/uss/Atropos, DBLP:conf/ccs/FuzzCache}. Existing techniques have made substantial progress in identifying individual vulnerability classes, including SQL injection, cross-site scripting, arbitrary file writes, and other taint-style flaws. However, vulnerabilities are not necessarily exploited in isolation. An individual flaw may provide only a restricted capability, such as reading a database record, modifying a configuration setting, or writing to a constrained file location. Although such a capability may appear insufficient to cause severe damage on its own, an attacker can combine it with other vulnerabilities or intended program behaviors to achieve a substantially more powerful effect. Consequently, assessing vulnerabilities independently may overlook attack paths through which seemingly limited capabilities lead to high-impact security consequences. This limitation also affects vulnerability prioritization based solely on individual severity scores~\cite{cvss_v4}, motivating chain-aware security analysis that considers how vulnerabilities interact~\cite{cisa2022ssvc}.

\begin{figure}[t]
    \centering
    \includegraphics[width=\columnwidth]{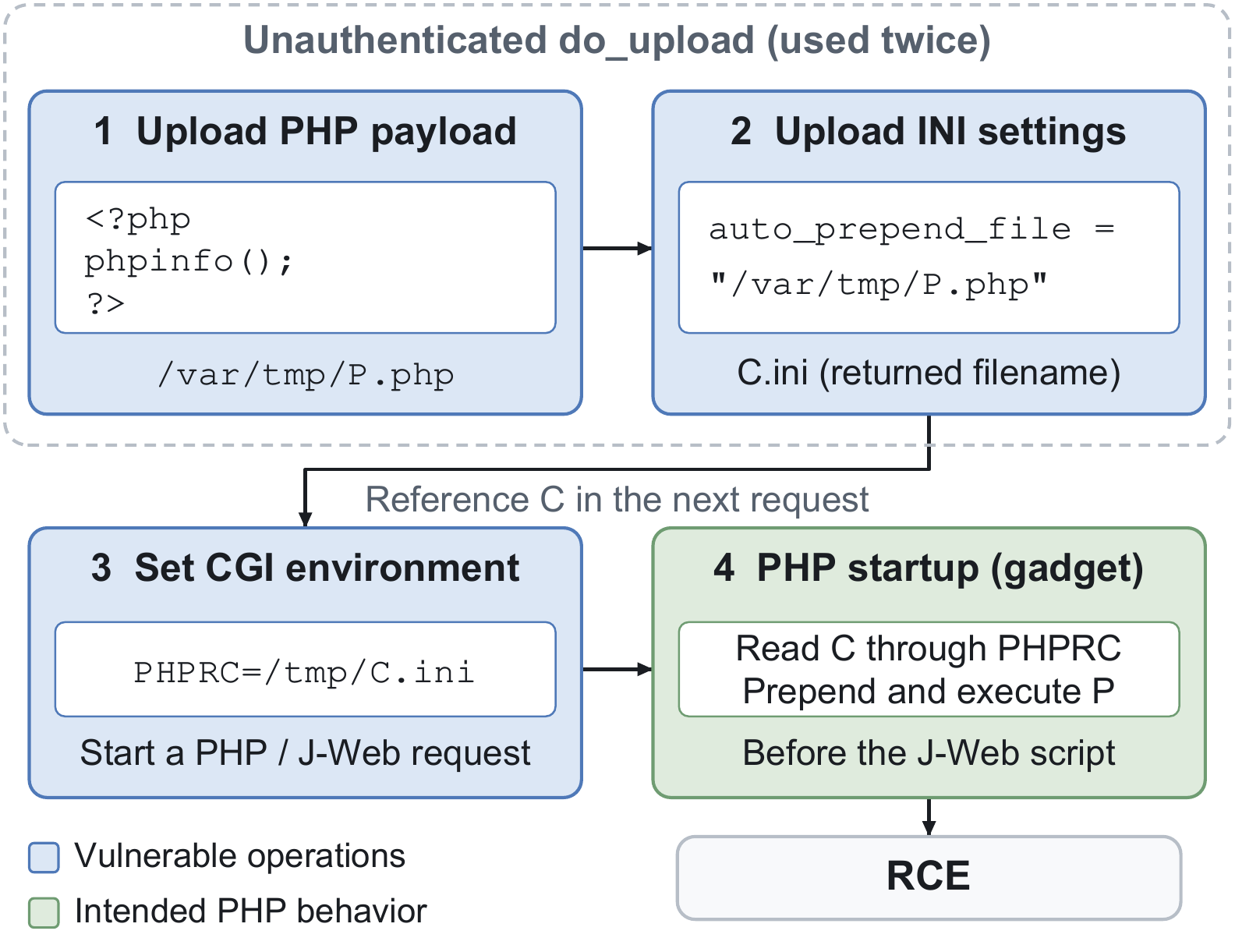}
    \caption{Juniper J-Web exploit chain. Two uses of one upload flaw (CVE-2023-36846~\cite{nvd_cve_2023_36846}) and environment injection (CVE-2023-36845~\cite{nvd_cve_2023_36845}) compose with PHP's normal startup behavior to achieve RCE. P and C abbreviate server-generated filenames.}
    \label{fig:juniper-chain}
\end{figure}

\Cref{fig:juniper-chain} illustrates this challenge through an exploit chain targeting Juniper's web interface~\cite{watchtowr_juniper_rce_2023}. An unauthenticated attacker first exploits an upload vulnerability~\cite{nvd_cve_2023_36846} to place a PHP payload in a temporary directory. The same vulnerability is then used to upload an INI configuration file whose \texttt{auto\_prepend\_file} directive references the payload. Next, an environment-injection vulnerability~\cite{nvd_cve_2023_36845} allows the attacker to set \texttt{PHPRC} to the configuration file. During the subsequent PHP CGI startup, the runtime loads the attacker-controlled configuration and executes the referenced PHP payload before processing the requested application script, ultimately achieving RCE. This case illustrates how individually limited vulnerabilities can be chained through their effects on shared security-relevant objects and their interactions with intended program behaviors.

Automated attack planning~\cite{DBLP:conf/uss/ChainReactor, DBLP:journals/corr/ALFA-Chains, DBLP:conf/uss/PentestGPT, DBLP:conf/asiaccs/PentestAgent, DBLP:conf/ijcai/PENTESTLLMAGENT} addresses a complementary aspect of this problem by reasoning about how attack actions can be composed to achieve security objectives. However, existing planning approaches typically assume a predefined set of attack components, rather than discovering previously unknown vulnerabilities and exploitable program behaviors from source code. Conversely, conventional vulnerability discovery identifies candidate flaws without using their potential roles in an exploit chain to guide further inspection. This separation between component discovery and attack planning leaves an important opportunity unexplored: the dependencies within an exploit chain can reveal what components are needed, and thus guide their discovery from source code.

This observation raises a central research question: \textbf{Can exploit-chain dependencies guide source-code analysis to jointly discover previously unknown vulnerabilities and exploitable program behaviors while constructing the chains that connect them?} Answering this question requires understanding what goals real-world exploit chains achieve, what components constitute them, and how these components are connected.

\subsection{Empirical Study}

Although vulnerability chaining is widely recognized as a security concern, existing vulnerability taxonomies~\cite{mitre_cwe, owasp-top10-2025} primarily characterize flaws individually, providing limited insight into how real-world exploit chains are composed. To bridge this gap, we conduct an empirical study of real-world web exploit chains, addressing three research questions:

\begin{itemize}
    \item \textbf{RQ1:} What high-impact security goals do real-world exploit chains ultimately achieve?
    \item \textbf{RQ2:} What behavioral components constitute these exploit chains?
    \item \textbf{RQ3:} What mechanisms connect adjacent components in exploit chains?
\end{itemize}

Our study yields three key findings. \ding{182} \textbf{Goal concentration.} Real-world exploit chains converge to a small set of high-impact goals, among which unauthorized code execution is prominent in our corpus. \ding{183} \textbf{Heterogeneous components.} Chains contain not only vulnerabilities, but also intended program behaviors that attackers repurpose to advance an attack. We call such security-relevant intended behaviors \emph{functional gadgets}. \ding{184} \textbf{Capability-based connections.} Vulnerabilities and functional gadgets are connected through security-relevant objects (e.g., files and databases) and the capabilities that are produced and consumed via these objects. Exploit-chain edges are therefore determined by whether one component's capabilities satisfy another component's preconditions, rather than by whether their vulnerability types match.

\subsection{Approach}

Based on these findings, we adopt a \textbf{precondition--capability abstraction} to represent both vulnerabilities and functional gadgets. A precondition specifies what an attacker needs to trigger a component, while a capability describes the operations an attacker can perform on security-relevant objects. Two components can be chained when the capabilities provided by one satisfy the preconditions of the other. Unlike traditional attack planning that assumes predefined components, our goal is to discover previously unknown vulnerabilities and functional gadgets from source code while constructing exploit chains. This requires inferring their preconditions and capabilities and using emerging dependencies to guide further discovery, raising two challenges:

\textbf{C1: How can we infer component preconditions and capabilities from source code?} Traditional static analyses recover low-level program semantics, such as control-flow constraints and data-flow operations applied to user input. Inferring preconditions and capabilities requires understanding the high-level security implications of these behaviors. For example, an analysis can determine that \texttt{updateOption(\$k,\$v)} is guarded by \texttt{canManage(\$u)} and that its arguments are user-controlled; recognizing the guard as an administrative precondition and determining the security consequences of the configuration change require application-level semantics.

\textbf{C2: How can we find capability producers and consumers in a target application?} The search space includes both vulnerabilities and numerous intended program behaviors. A precondition alone does not indicate which vulnerability or functional gadget could satisfy it, while a capability does not reveal which functional gadget could consume or amplify it. Without explicit guidance, an LLM lacks a focused search target and must explore a prohibitively large space of possible components.

To address these challenges, we design and implement \tool, a goal-directed LLM-agent framework for exploit-chain discovery in PHP applications. Given an application's source code and a target security goal, \tool jointly discovers vulnerabilities and functional gadgets while constructing exploit chains, without requiring a predefined vulnerability list. \tool realizes this discovery process through three complementary mechanisms. \ding{182} \textbf{LLM-based property inference} leverages Large Language Models (LLMs) to bridge low-level program behaviors and high-level security semantics, inferring component preconditions and capabilities from source code. \ding{183} \textbf{Dependency-driven exploration} uses unmet preconditions to guide backward searches for capability producers and discovered capabilities to guide forward searches for functional gadgets that consume or amplify them, interleaving component discovery with exploit-chain construction. \ding{184} \textbf{Object-centered chaining knowledge} distills our empirical findings into structured knowledge that maps security-relevant objects and capabilities to potential vulnerabilities and functional gadgets, guiding source-code inspection toward promising targets.

Driven by these mechanisms, \tool operates in an iterative, goal-directed exploration loop. Starting from a target security goal, \tool maintains a worklist of discovery tasks derived from unmet preconditions and discovered capabilities. Specialized agents inspect candidate code regions to identify vulnerabilities and functional gadgets, whose inferred preconditions and capabilities are used to incrementally construct an exploit graph. Each newly discovered component updates the graph and generates further exploration tasks, allowing component discovery and exploit-chain construction to reinforce each other. Finally, \tool synthesizes goal-reaching exploit chains from the constructed graph.

\subsection{Contributions}

We evaluate \tool against Codex~\cite{openai_codex}, a state-of-the-art general-purpose coding agent, on 15 PHP applications across three shared LLM backends. \tool achieves 42.02\%--44.63\% recall for components on paths to the security goal and recovers up to $4.4\times$ as many correct conditional goal-reaching edges as Codex (185 vs.\ 42 under DeepSeek-v4-Flash). Five ablation studies further demonstrate how explicit precondition--capability modeling, functional gadgets, and empirical chaining knowledge support component discovery and graph construction, and how joint discovery identifies chain-relevant vulnerabilities missed by discovery-then-plan. We additionally evaluate 30 real-world PHP applications, documenting 20 vulnerabilities across seven applications and 13 exploit chains across seven applications. Nine chains involve functional gadgets, and nine achieve RCE. These results show that guiding code inspection by capability dependencies helps agents identify and connect components that contribute to high-impact attack paths.

This paper makes the following contributions:
\begin{itemize}
    \item \textbf{Empirical Characterization of Web Exploit Chains.} We conduct a systematic study of 386 real-world web exploit chains, characterizing their security goals, the roles of vulnerabilities and functional gadgets, and the security-relevant object dependencies that connect adjacent components.

    \item \textbf{Goal-directed Joint Discovery and Composition.} We develop \tool, an LLM-agent framework that jointly discovers vulnerabilities and functional gadgets while constructing their capability dependencies. \tool combines backward precondition-guided discovery, forward capability-guided gadget exploration, and object-centered chaining knowledge derived from empirical exploit-chain analysis.

    \item \textbf{Systematic Evaluation and Real-world Findings.} We evaluate \tool against Codex on 15 PHP applications across multiple LLM backends and investigate its core design choices through five ablation studies. An additional evaluation of 30 real-world applications documents 20 vulnerabilities and 13 exploit chains, including 9 that achieve RCE.

\end{itemize}

\section{Empirical Study}
\label{sec:empirical-study}

In this section, we first describe the data collection and analysis procedure for our empirical study, and then systematically analyze real-world web exploit chains to answer RQ1--RQ3 introduced in \Cref{sec:introduction}.

\subsection{Data Collection and Analysis}
\label{sec:study-data}

\noindent\textbf{Data Collection.} We collect public exploit artifacts from three widely used repositories: Metasploit Framework, nuclei-templates, and Awesome-POC. Because our study focuses on web exploit chains, we retain only HTTP-oriented exploit modules. As summarized in \Cref{tab:study-dataset}, this process yields 7,251 candidate artifacts.

\begin{table}[h]
\centering
\caption{Summary of the collected exploit artifacts and retained exploit chains.}
\label{tab:study-dataset}
\small
\begin{tabular}{lcc}
    \toprule
    \textbf{Repository} & \textbf{Artifacts} & \textbf{Exploit Chains} \\
    \midrule
    Metasploit Framework & 1144 & 205 \\
    Nuclei Templates & 4999 & 114 \\
    Awesome POC & 1108 & 67 \\
    \midrule
    \textbf{Total} & \textbf{7251} & \textbf{386} \\
    \bottomrule
\end{tabular}
\end{table}

\noindent\textbf{Artifact Analysis and Screening.} For each candidate artifact, we first use DeepSeek-V4-Pro~\cite{deepseekv4} to identify the involved vulnerability classes, decompose the exploitation procedure into ordered steps, and summarize the dependencies among these steps. We then manually validate and, when necessary, correct the model output against the original artifact. We retain an artifact as an exploit chain only when it contains multiple dependent exploitation steps, such that an earlier step enables a subsequent one. We exclude artifacts that describe only an isolated vulnerability or involve non-web vulnerability classes, such as buffer overflows. This process results in 386 exploit chains for subsequent analysis.

\subsection{RQ1: Security Goals of Exploit Chains}
\label{sec:rq1}

To characterize the ultimate goals of real-world exploit chains, we classify the final impact of each chain according to the CWE Enumeration of Technical Impacts~\cite{cwe_technical_impacts}. This taxonomy categorizes an attack by its resulting security consequence, including executing unauthorized code or commands, reading data, gaining privileges or assuming an identity, and modifying data. We label each chain by its final impact only; impacts produced by intermediate steps are not counted separately.

\begin{table}[t]
\centering
\caption{RQ1: Final technical impacts of the collected exploit chains.}
\label{tab:rq1-goals}
\small
\begin{tabular}{@{}p{0.58\columnwidth}rr@{}}
    \toprule
    \textbf{Technical Impact} & \textbf{Count} & \textbf{Ratio} \\
    \midrule
    Execute Unauthorized Code or Commands & 366 & 94.8\% \\
    Read Data & 14 & 3.6\% \\
    Gain Privileges or Assume Identity & 5 & 1.3\% \\
    Modify Data & 1 & 0.3\% \\
    \midrule
    \textbf{Total} & \textbf{386} & \textbf{100.0\%} \\
    \bottomrule
\end{tabular}
\end{table}

\Cref{tab:rq1-goals} shows a highly concentrated distribution. Of the 386 exploit chains, 366 (94.8\%) ultimately execute unauthorized code or commands. Only 14 chains (3.6\%) terminate in data disclosure, five (1.3\%) in privilege gain or identity assumption, and one (0.3\%) in data modification. Thus, the remaining three impact categories together account for only 5.2\% of the dataset.

\begin{rqbox}
\textbf{Finding 1.} Real-world web exploit chains overwhelmingly converge on unauthorized code or command execution, indicating that intermediate vulnerabilities are most often composed to achieve this high-impact goal.
\end{rqbox}

\subsection{RQ2: Components of Exploit Chains}
\label{sec:rq2}

To identify what constitutes an exploit chain, we decompose each chain into security-relevant program behaviors that advance the exploitation process. We first code each behavior according to its role in the chain without imposing a predefined node taxonomy. The resulting codes reveal two recurring categories. A \emph{vulnerability component} exploits unintended program behavior to obtain a capability, whereas a \emph{functional gadget} repurposes an intended program behavior to consume or amplify a capability obtained by preceding components. We then apply these definitions to annotate all components in the dataset.

\begin{table}[t]
    \centering
    \caption{RQ2: Composition of the collected exploit chains by data source.}
    \label{tab:rq2-components}
    \small
    \setlength{\tabcolsep}{3pt}
    \begin{tabular}{@{}lrrrr@{}}
        \toprule
        \textbf{Repository} & \textbf{Chains} & \textbf{Vuln.} & \textbf{Gadgets} & \textbf{Gadget Chains} \\
        \midrule
        Metasploit & 205 & 434 & 39 & 39 (19.0\%) \\
        Nuclei & 114 & 237 & 3 & 3 (2.6\%) \\
        Awesome-POC & 67 & 137 & 7 & 7 (10.4\%) \\
        \midrule
        \textbf{Total} & \textbf{386} & \textbf{808} & \textbf{49} & \textbf{49 (12.7\%)} \\
        \bottomrule
    \end{tabular}
\end{table}

\Cref{tab:rq2-components} summarizes the resulting component distribution. The 386 exploit chains contain 857 components, including 808 vulnerability components (94.3\%) and 49 functional gadgets (5.7\%). Overall, 49 chains (12.7\%) contain a functional gadget. Gadget-containing chains are most frequent in Metasploit Framework (19.0\%), followed by Awesome-POC (10.4\%) and nuclei-templates (2.6\%). These results show that exploit chains are not composed exclusively of vulnerabilities: attackers also reuse intended program behaviors as intermediate exploitation steps.

\begin{figure}[t]
    \centering
    \includegraphics[width=\columnwidth]{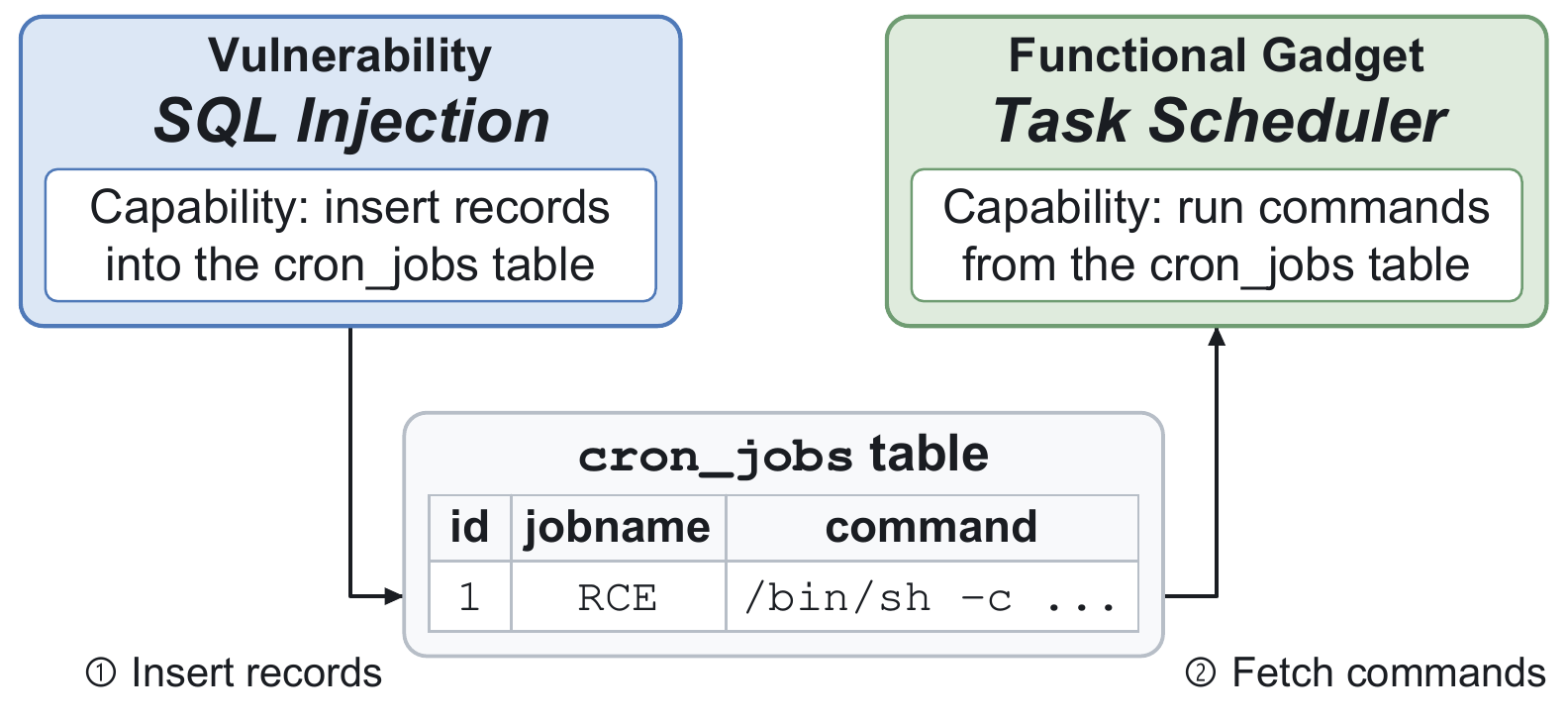}
    \caption{An exploit chain in FreePBX~\cite{rapid7_freepbx_unauth_sqli_rce}, where an unauthenticated SQL injection leverages an intended task scheduler gadget to achieve RCE.}
    \label{fig:gadget-example}
\end{figure}

\Cref{fig:gadget-example} illustrates how a functional gadget bridges an isolated flaw to RCE in FreePBX (CVE-2025-57819) from Metasploit Framework ~\cite{rapid7_freepbx_unauth_sqli_rce}. \ding{182} An attacker exploits an unauthenticated SQL injection to insert a malicious command record into the \texttt{cron\_jobs} table. \ding{183} While the injection alone cannot execute OS commands, FreePBX's intended task scheduler periodically fetches and executes entries from this table. By consuming the state modified by the vulnerability, this benign gadget elevates an isolated database write into RCE.

\begin{rqbox}
\textbf{Finding 2.} Real-world web exploit chains comprise both vulnerabilities and functional gadgets. Functional gadgets appear in 12.7\% of the studied chains, showing that intended program behaviors can serve as chain components to extend previously obtained capabilities.
\end{rqbox}

\subsection{RQ3: Mechanisms Connecting Adjacent Components}
\label{sec:rq3}

To understand how exploit chains are constructed, we investigate the underlying mechanisms that allow two adjacent components to connect. We first examine whether exploit chains follow fixed vulnerability-type patterns, and then analyze the semantic object-level dependencies that actually enable component composition.

\noindent\textbf{Vulnerability-Type Composition.}
A natural hypothesis is that exploit chains follow predetermined transitions between fixed vulnerability types. To evaluate this, we analyze the co-occurrence of adjacent vulnerability types across the 386 exploit chains. Appendix \Cref{fig:vuln-heatmap} presents the co-occurrence heatmap for the top-15 most frequent CWEs in our dataset.

The matrix reveals an expansive, many-to-many composition landscape rather than rigid pairings. Crucially, a single vulnerability type often yields different capabilities depending on how it is exploited. 
For instance, the heatmap shows that SQL Injection (CWE-89) frequently co-occurs with both authentication flaws (CWE-287, CWE-288) and file-handling flaws (CWE-434, CWE-22). In practice, an attacker can leverage SQLi to leak credentials to bypass authentication, or alternatively use it to write files to disk (e.g., via \texttt{INTO OUTFILE}) to satisfy downstream execution preconditions. 
Similarly, Path Traversal (CWE-22) pairs with over ten distinct CWEs, while Command Injection (CWE-78) is reachable via diverse precursors ranging from deserialization (CWE-502) to hard-coded credentials (CWE-798). 
These patterns demonstrate that \emph{syntactic vulnerability types alone do not dictate composability}, prompting us to examine the underlying mechanisms that enable these pairings.

\begin{table*}[t]
    \centering
    \caption{RQ3: Security domains connecting adjacent components.}
    \label{tab:rq3-domains}
    \small
    \setlength{\tabcolsep}{6pt}
    \begin{tabular}{@{}lr@{\hspace{18pt}}lr@{\hspace{18pt}}lr@{}}
        \toprule
        \multicolumn{2}{c}{\textit{Vuln. $\rightarrow$ Vuln.}} &
        \multicolumn{2}{c}{\textit{Vuln. $\rightarrow$ Gadget}} &
        \multicolumn{2}{c}{\textit{Gadget $\rightarrow$ Vuln.}} \\
        \cmidrule(r){1-2}\cmidrule(lr){3-4}\cmidrule(l){5-6}
        \textbf{Domain} & \textbf{Count (\%)} &
        \textbf{Domain} & \textbf{Count (\%)} &
        \textbf{Domain} & \textbf{Count (\%)} \\
        Credential & 190 (45.0) &
        File System (App. Configuration) & 29 (59.2) &
        File System (Logs) & 4 (57.1) \\
        File System & 83 (19.7) &
        Task Scheduler & 8 (16.3) &
        File System (Images) & 1 (14.3) \\
        Network Reachability & 62 (14.7) &
        App. Package & 7 (14.3) &
        File System (Scripts) & 1 (14.3) \\
        Runtime Memory & 60 (14.2) &
        Database/Runtime Memory & 3 (6.1) &
        Credential & 1 (14.3) \\
        Database & 27 (6.4) &
        Credential & 2 (4.1) &
        & \\
        \bottomrule
    \end{tabular}
\end{table*}

\noindent\textbf{Object-Driven Connection Mechanisms.}
Because vulnerability types do not explain composability, we inspect each component pair $(u, v)$ from a semantic perspective to uncover what actually bridges them. We observe that adjacent components are connected through \emph{security-relevant objects}: component $u$ provides a capability to create, modify, or access an object, which satisfies the security state or access required by component $v$ (its precondition).

To systematically annotate and classify these objects without subjective bias, we ground our categorization in the MITRE D3FEND Artifact~\cite{mitre_d3fend_dao}. In D3FEND, an \emph{Artifact} represents a fundamental digital entity (e.g., credentials, files, or network traffic) that system defenses monitor and protect. Following this standard, we label the specific artifact affected by each exploit edge and group similar artifacts into normalized security domains. For example, passwords, API keys, and session tokens map to the \emph{Credential} domain, while configuration files, log files, and upload paths map to the \emph{File System} domain.

Across the dataset, the 857 components form 478 adjacent connections. As summarized in \Cref{tab:rq3-domains}, these connections span three interaction paradigms, each relying on distinctly different security domains.
\begin{itemize}
    \item \textbf{\textit{Vuln. $\rightarrow$ Vuln.} (422 connections):} Dominated by \emph{Credentials} (45.0\%) and \emph{File System} (19.7\%). In these cases, an initial vulnerability (e.g., SQLi or sensitive data disclosure) is typically used to extract administrative credentials, which then satisfies the authentication precondition of a subsequent privileged vulnerability.
    \item \textbf{\textit{Vuln. $\rightarrow$ Gadget} (49 connections):} Heavily driven by application-managed state, especially \emph{App. Configuration} (59.2\%) and \emph{Task Schedulers} (16.3\%). Here, an attacker exploits a vulnerability to alter a persistent configuration object, which is subsequently processed by benign application logic to trigger code execution.
    \item \textbf{\textit{Gadget $\rightarrow$ Vuln.} (7 connections):} Primarily mediated through the file system (e.g., \emph{Logs} accounting for 57.1\%). An attacker uses legitimate logging functionality to write payloads to disk, establishing the file precondition required for a downstream vulnerability (e.g., local file inclusion).
\end{itemize}

\begin{rqbox}
\textbf{Finding 3.} Adjacent components are not governed by fixed vulnerability types. Instead, their composition is driven by dependencies on security-relevant objects, where the capabilities produced by one component satisfy the preconditions required by the next.
\end{rqbox}

In \Cref{sec:worklist-search}, we turn these object-level dependencies into knowledge that guides searches for capability producers and consumers.

\section{Methodology}
\label{sec:methodology}

The three empirical findings guide the design of \tool. Finding~1 shows that chains converge on a small set of high-impact outcomes, motivating a goal-directed search that starts from a target such as code execution and works backward through its required capabilities. Finding~2 motivates searching for both vulnerabilities and functional gadgets, since intended program behaviors can supply essential intermediate capabilities. Finding~3 motivates a unified precondition--capability model that connects components through capabilities over security-relevant objects. These object-level dependencies also form the basis of our chaining knowledge, which directs the search toward likely capability producers and consumers.

\begin{figure*}[t]
    \centering
    \includegraphics[width=0.95\textwidth,pagebox=cropbox]{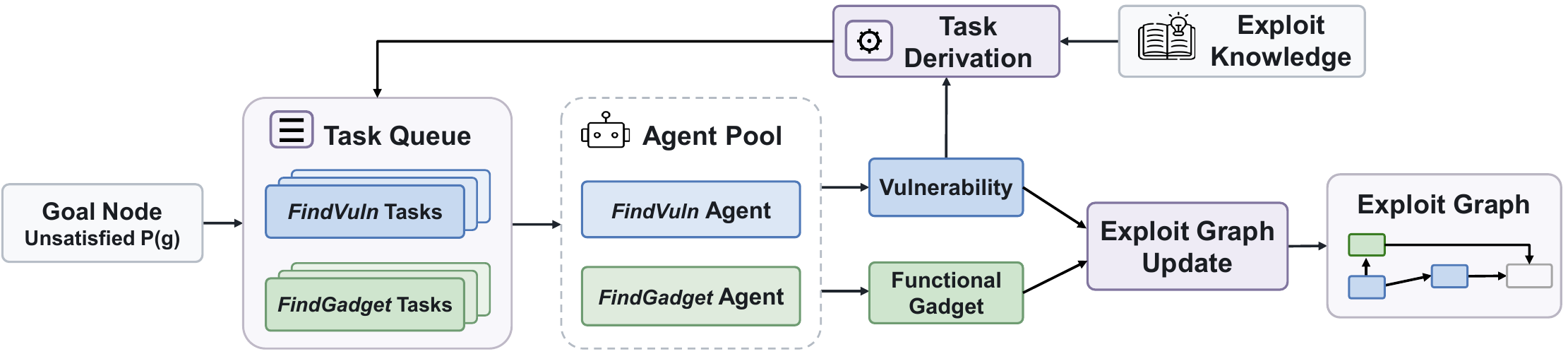}
    \caption{Workflow of \tool. Goal-guided search iteratively derives vulnerability- and gadget-search tasks, discovers components, and updates the ExpGraph using empirical chaining knowledge.}
    \label{fig:workflow}
\end{figure*}

Figure~\ref{fig:workflow} presents the workflow of \tool. Given the target PHP application's source code and a security goal, \tool iteratively discovers vulnerabilities and functional gadgets and records their semantic connections in an Exploit Graph (ExpGraph) defined in~\Cref{sec:exploit-graph}. The initial graph contains only a virtual goal node; vulnerability and gadget nodes are discovered from the application source code during exploration. \tool maintains a worklist of precondition-guided searches for vulnerabilities and gadgets, together with capability-guided gadget searches. These searches are guided by the chaining knowledge derived from Finding~3. Each discovered component is modeled by its preconditions and capabilities and then added to the graph. After the worklist is exhausted, \tool synthesizes goal-reaching exploit chains from the graph.

\subsection{Exploit Graph}
\label{sec:exploit-graph}

The ExpGraph records reusable exploit-chain components and their precondition--capability dependencies. \tool incrementally updates it during discovery and uses it to synthesize exploit chains after the search terminates.

\noindent\textbf{Graph Definition.} Let an ExpGraph be a directed graph $\mathit{G}=(N,L)$, where $N$ is the set of component nodes and $L\subseteq N\times N$ is the set of directed links.

\noindent\textbf{Nodes.} Each node $n\in N$ has attributes $\langle\tau(n),P(n),C(n)\rangle$: its type, precondition set, and capability set provided to the attacker by triggering the component. The type $\tau(n)\in\{\mathsf{Vuln},\mathsf{Gadget},\mathsf{Goal}\}$ distinguishes a discovered vulnerability, an intended program behavior reusable in an exploit chain, and the virtual goal node $g$, respectively. For $g$, $P(g)$ specifies the target capability.

We represent preconditions and capabilities using the predicate structure $\mathsf{Domain.Action}(a_1,\ldots,a_k)$, where the domain denotes a security-relevant object, the action specifies an operation, and the arguments capture targets and constraints. All fields are expressed in natural language, allowing application-specific properties to be described without a fixed predicate vocabulary.

\noindent\textbf{Links.} A directed link $\ell=(u,v)\in L$ exists when at least one capability of $u$ semantically satisfies a precondition of $v$. Each link is labeled with its matching set
\begin{equation}
    M(\ell)=\{(c,p)\mid c\in C(u),\ p\in P(v),\ c\models p\},
\end{equation}
where $c\models p$ denotes that capability $c$ semantically satisfies precondition $p$. Thus, $\ell$ exists iff $M(\ell)\neq\emptyset$. These matches are used for chain synthesis.

\subsection{Knowledge-Guided Discovery and Graph Construction}
\label{sec:worklist-search}

\noindent\textbf{Object-Centered Chaining Knowledge.} Following Finding~3, we organize the annotated adjacent-component pairs by their shared security objects and transferred capabilities. For each group, we summarize the vulnerability classes and intended operations that produce or consume those capabilities. Each knowledge entry records a search direction, an object domain, a required or produced capability, candidate component families, and search hints such as relevant sink families or object reads and writes. \Cref{tab:chaining-knowledge} shows representative entries.

\begin{table}[t]
    \centering
    \caption{Examples of object-centered chaining knowledge. $V$ denotes a vulnerability and $G$ a functional gadget.}
    \label{tab:chaining-knowledge}
    \footnotesize
    \setlength{\tabcolsep}{3pt}
    \renewcommand{\arraystretch}{1.1}
    \begin{tabular}{@{}p{0.13\columnwidth}p{0.30\columnwidth}p{0.50\columnwidth}@{}}
        \toprule
        \textbf{Query} & \textbf{Object / capability} & \textbf{Recommended search} \\
        \midrule
        $P \rightarrow V$ & Credential: obtain credentials & SQL injection or information disclosure; inspect credential stores and related read paths. \\
        \addlinespace
        $P \rightarrow G$ & File system: create a file with controlled content & Logging routines; inspect writes of request-controlled content. \\
        \addlinespace
        $C \rightarrow G$ & Configuration: modify a setting & Configuration loaders; follow reads and uses of the modified setting. \\
        \bottomrule
    \end{tabular}
\end{table}

For each search task, the agent identifies the security object and the required or provided capability specified by the query, then consults the matching entries. A precondition guides the search toward capability producers, including both vulnerabilities and functional gadgets; a provided capability guides it toward gadget consumers. The recommendations narrow inspection to promising component families, relevant object accesses, and enclosing features. Concrete object bindings and capability constraints are then checked during property inference and graph update, rather than inferred from component types alone.

\begin{algorithm}[t]
\small
\DontPrintSemicolon
\caption{Goal-guided ExpGraph construction}
\label{alg:worklist}
\KwIn{target PHP application $\mathcal{A}$ and target security goal $g$}
\KwOut{Exploit Graph $\mathcal{G}$}
$\mathcal{G}\leftarrow(\{g\},\emptyset)$; $Seen\leftarrow\emptyset$\;
$Q\leftarrow\{\mathsf{FindVuln}(p,g)\mid p\in P(g)\}$\;
$Q\leftarrow Q\cup\{\mathsf{FindGadget}^{P}(p,g)\mid p\in P(g)\}$\;
\While{$Q\neq\emptyset$}{
    $t\leftarrow Q.\mathsf{pop}()$\;
    \If{$key(t)\notin Seen$}{
        $Seen\leftarrow Seen\cup\{key(t)\}$\;
        \ForEach{$b\in\mathsf{RunTask}(t,\mathcal{A})$}{
            $n\leftarrow\mathsf{InferProperties}(b)$\;
            $\mathsf{MatchAndLink}(n,\mathcal{G})$\;
            $Q\leftarrow Q\cup\{\mathsf{FindVuln}(p,n)\mid p\in P(n)\}$\;
            $Q\leftarrow Q\cup\{\mathsf{FindGadget}^{P}(p,n)\mid p\in P(n)\}$\;
            $Q\leftarrow Q\cup\{\mathsf{FindGadget}^{C}(c,n)\mid c\in C(n)\}$\;
        }
    }
}
\textbf{return} $\mathcal{G}$\;
\end{algorithm}

Algorithm~\ref{alg:worklist} summarizes the goal-guided search. The worklist maintains two task types:
\begin{itemize}
    \item A $\mathsf{FindVuln}(p,n_t)$ task searches for a vulnerability whose capabilities satisfy precondition $p$ of target node $n_t$.
    \item A gadget-search task has two modes: $\mathsf{FindGadget}^{P}(p,n_t)$ searches for a functional gadget whose capabilities satisfy precondition $p$ of target node $n_t$, while $\mathsf{FindGadget}^{C}(c,n_s)$ searches for one that consumes or amplifies capability $c$ of source node $n_s$.
\end{itemize}

The two task types are handled by \textsc{FindVulnAgent} and \textsc{FindGadgetAgent}, respectively. Both agents use the object-centered knowledge above to focus their source-code searches. Each task is keyed by its type, normalized property, anchor node, and, for gadget searches, its mode. \tool processes each key once to avoid repeated searches and cycles.

\tool initializes the ExpGraph with the virtual goal node and creates both vulnerability- and gadget-search tasks for each goal precondition. Each discovered component is modeled, inserted into the graph, and matched against all existing nodes. Its preconditions then generate both types of search tasks, while its capabilities generate capability-guided gadget-search tasks. The algorithm terminates when no unseen task remains and returns the constructed ExpGraph.

\begin{figure}[t]
    \centering
    \includegraphics[width=\columnwidth,pagebox=cropbox]{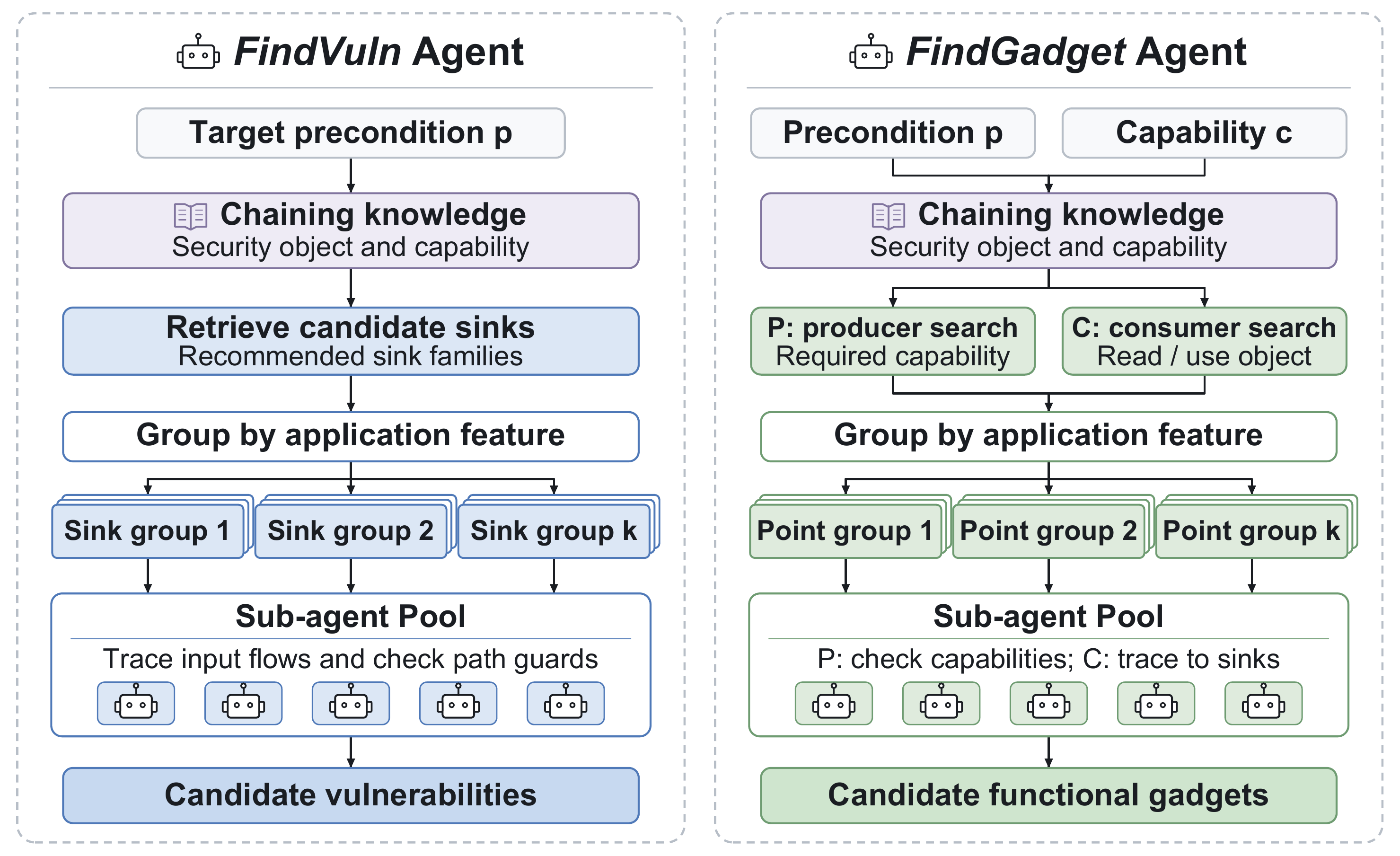}
    \caption{Discovery workflows of \textsc{FindVulnAgent} (left) and \textsc{FindGadgetAgent} (right). Chaining knowledge guides sink or object-access retrieval. Candidates are grouped by application feature and dispatched to sub-agent pools for focused source-code inspection.}
    \label{fig:agent-workflow}
\end{figure}

\subsubsection{Precondition-Guided Vulnerability Discovery}

The left side of Figure~\ref{fig:agent-workflow} shows how \textsc{FindVulnAgent} processes a precondition $p$ of target node $n_t$. The agent identifies the security object and required capability in $p$, then queries the chaining knowledge for vulnerability classes and sink families that may produce that capability. It retrieves candidate sinks from the target application and groups them by their enclosing feature and functional similarity. This grouping keeps related sinks, their shared guards, and calling contexts within one inspection scope, reducing the number of sub-agents required and improving analysis efficiency. The resulting sink groups are dispatched to a sub-agent pool for parallel analysis.

Within each group, a sub-agent traces input flows to the relevant sinks and examines the path guards. It checks whether attacker-controlled data can reach the operation under the enclosing authentication, permission, and branch conditions, and whether intervening validation prevents the required capability. A candidate vulnerability is retained when the inspected code supports a capability that can establish $p$. The agent returns these candidates for property inference and graph update.

\subsubsection{Precondition- and Capability-Guided Gadget Discovery}

The right side of Figure~\ref{fig:agent-workflow} shows the two search modes of \textsc{FindGadgetAgent}. In the precondition-guided (P) mode, the agent identifies the security object and capability required by $p$ of target node $n_t$. It consults the chaining knowledge to retrieve intended operations that may produce this capability. For example, ordinary logging may establish a file-inclusion precondition by writing attacker-controlled content to an includable file.

In the capability-guided (C) mode, the agent identifies the security object and available operation described by capability $c$ of source node $n_s$, then queries the knowledge for likely consumer patterns. It retrieves program points that read or otherwise use that object. The search focuses on accesses to the relevant state, such as reads of a modified configuration setting or database record.

In both modes, candidate producer or consumer points are grouped by application feature and dispatched to a sub-agent pool for the same efficiency reasons as in vulnerability discovery. P-mode inspections check whether the intended operation produces the required capability under compatible object and value constraints. C-mode inspections trace forward from each consumer through downstream data and state dependencies toward sensitive operations, checking how the affected state influences those operations. A reported gadget must establish the requested precondition or consume the source capability to produce a capability relevant to the target goal. The retained candidates then undergo the same property inference and graph update as vulnerabilities.

\subsubsection{Property Inference and Graph Update}

After either discovery agent reports a candidate component, the LLM examines its source code and relevant calling contexts. It identifies attacker-controlled inputs and traces their propagation through assignments, calls, and accesses to persistent state. It checks sanitization and validation along the flow to determine what control reaches the relevant operation. It also identifies the path conditions required to reach that operation, including authentication, permissions, configuration settings, and branch conditions.

The LLM then reconstructs the sink arguments as symbolic expressions in terms of program inputs and intermediate definitions, retaining transformations and fixed path or value constraints. It uses these expressions and path conditions to infer $P(n)$ and $C(n)$: required privileges and object access become preconditions, while attacker-controllable operations on security-relevant objects determine the provided capabilities. For example, an administrative guard becomes a permission precondition, whereas a fixed directory or filename suffix restricts a file-write capability. The model also considers whether the program behavior is intended when distinguishing vulnerabilities from gadgets.

\tool then inserts $n$ into the ExpGraph and compares it with every existing node $x\in N\setminus\{n\}$. For each prospective link, the LLM semantically compares these descriptions, accommodating wording differences across agents while checking whether a capability satisfies a precondition on the same object under compatible path and value constraints. It adds $(n,x)$ when $M(n,x)\neq\emptyset$ and $(x,n)$ when $M(x,n)\neq\emptyset$, recording the corresponding matching set on each edge. Independently of the resulting matches, each precondition generates vulnerability- and gadget-search tasks, while each capability generates a capability-guided gadget-search task.

\subsection{Exploit-Chain Synthesis}
\label{sec:chain-synthesis}

After exploration, a chain-synthesis agent extracts goal-reaching exploit chains from the ExpGraph. Enumerating all combinations of alternative components can lead to path explosion. To control this growth and provide human analysts with concise, auditable exploit chains, the agent reduces redundant alternatives during backward traversal as follows.
\ding{182} \textbf{Backward expansion.} Starting from the virtual goal node, the agent uses the matching sets on incoming edges to identify predecessors that establish each required precondition. It combines multiple predecessors when the required capabilities come from different sources, checking that their object bindings and path constraints are compatible.
\ding{183} \textbf{Grouping alternatives.} For the same capability requirements, the agent groups candidate predecessors that provide equivalent capabilities on the same security objects under compatible execution contexts. For example, multiple privilege-escalation vulnerabilities may each provide the administrator privilege required by a downstream component. Grouping is used during synthesis; individual component nodes remain in the ExpGraph.
\ding{184} \textbf{Selecting representatives.} Before choosing a representative, the agent examines each alternative's upstream dependencies and compares the entry assumptions left after other components establish its preconditions. An alternative is omitted when another provides the same required capabilities with no stronger entry assumptions. Alternatives with incomparable assumptions are retained; when assumptions are equivalent, the agent prefers the alternative with fewer components.
\ding{185} \textbf{Ordering and output.} The agent orders the components of each retained chain according to their dependencies and records the matched capability--precondition pairs for every transition. Preconditions not established within the chain are reported as explicit entry assumptions.

\section{Evaluation}
\label{sec:evaluation}

We evaluate \tool by answering the following research questions:
\begin{itemize}
    \item \textbf{RQ4: Comparison to State-of-the-Art.} How effectively does \tool discover exploit chains compared with state-of-the-art coding agents?
    \item \textbf{RQ5: Ablation Studies.} How does each core component contribute to \tool's exploit-chain discovery performance?
    \item \textbf{RQ6: Real-World Applications.} Can \tool discover previously unknown vulnerabilities and compose them into high-impact exploit chains in real-world PHP applications?
\end{itemize}

\subsection{Experimental Setup}

\noindent\textbf{Compared Tools.} To evaluate the effectiveness of \tool, we compare it against Codex~\cite{openai_codex} (version v0.147.0), an open-source, state-of-the-art general-purpose coding agent.

\noindent\textbf{Evaluation Benchmarks.} We evaluate \tool and Codex on a benchmark of 15 real-world PHP web applications adopted from \textsc{Zipper}~\cite{DBLP:conf/uss/ZIPPER}, which have been widely studied in prior PHP vulnerability discovery research~\cite{alhuzali2018navex, luo2022tchecker, RIPS-A, al2022testability, DBLP:conf/uss/ZIPPER}. These applications cover diverse domains, such as CMS (e.g., WordPress and Joomla), e-commerce (e.g., oscommerce2), and management tools, with codebases ranging from 3.9K to 273.8K lines of code. This diverse suite exhibits varied architectural complexities and rich attack surfaces, providing a realistic testbed for exploit chain discovery.

\noindent\textbf{Configurations.}
We evaluate both \tool and Codex across three representative LLMs: \textit{DeepSeek-v4-Flash-0731}~\cite{deepseek_v4_flash}, \textit{GLM-5.3-Flash}~\cite{glm_5_3_flash}, and \textit{Qwen-3.8-Flash}~\cite{qwen_3_8_flash}, as well as Codex's officially recommended model, \textit{GPT-5.4}~\cite{openai_gpt5_4}. All models are configured with their maximum reasoning effort. Experiments were conducted on a machine equipped with an Intel Core i7-11800H CPU and 64GB RAM running 64-bit Windows 11.
For RQ4 and RQ5, we set the security goal to RCE to facilitate evaluation. For RQ6, we use all categories in the CWE Enumeration of Technical Impacts~\cite{cwe_technical_impacts} as security goals.

\noindent\textbf{Evaluation Metrics.}
We evaluate \tool across two complementary dimensions: individual component discovery and goal-directed exploit graph construction. For component discovery, we report True Positives (TP), False Positives (FP), and Precision for both vulnerabilities and functional gadgets.

To evaluate exploit graph construction, we focus on the components and transitions that actively contribute to the specified security goal. Formally, let $G=(V, E)$ denote the ground-truth exploit graph, where nodes represent vulnerabilities and gadgets, and edges denote capability transfers. We extract the \emph{goal-reaching subgraph} $G_g=(V_g, E_g)$, where $V_g \subseteq V$ comprises all nodes possessing a directed path to a goal node, and $E_g \subseteq E$ contains the connecting edges along these paths. We report the TP, FN, and Recall over $V_g$ as \textbf{Goal Reach}, measuring the tool's ability to discover components critical to reaching the goal.

We also report TP, FP, and precision as \textbf{Edge}, considering only distinct directed edges whose two endpoints are true-positive vulnerabilities or gadgets matched to ground-truth nodes. Edges incident to goal nodes are excluded. An eligible edge is a TP if its endpoint pair belongs to $E$, and an FP otherwise. This metric covers all eligible component edges, including those outside $G_g$.

Furthermore, to evaluate edge-construction accuracy without conflating it with missed nodes, we define the \textbf{Conditional Goal-Reaching Edge}. Given the set of correctly identified nodes in a run $V_{\mathrm{TP}} \subseteq V_g$, we restrict the candidate ground-truth edges to those whose endpoints were both successfully discovered:
\begin{equation*}
    E_{\mathrm{cond}} = \{(u, v) \in E_g \mid u, v \in V_{\mathrm{TP}}\}
\end{equation*}
Let $E_{\mathrm{pred}}$ denote the edges predicted by the tool. We measure $\mathrm{TP} = |E_{\mathrm{pred}} \cap E_{\mathrm{cond}}|$ and compute the conditional recall as $\mathrm{TP} / |E_{\mathrm{cond}}|$. This metric isolates the agent's semantic reasoning capability in connecting components from its ability to spot individual bugs, directly assessing whether the tool can correctly infer the dependencies between discovered components.

\noindent\textbf{Ground-Truth.}
To construct the ground-truth exploit graph $G=(V,E)$, we combine automated PoC synthesis with manual verification. We first aggregate all components and transitions reported across all runs of \tool and Codex, alongside documented benchmark vulnerabilities~\cite{DBLP:conf/uss/ZIPPER}. We then task Codex (powered by \textit{DeepSeek-v4-Flash-0731}) with generating Proof-of-Concept (PoC) scripts for each candidate. For candidates validated by Codex, we manually execute the PoCs to confirm their security impact. For candidates that Codex failed to exploit, two authors independently audit the source code to verify any valid findings missed by the model. Finally, edges are established only after confirming that the predecessor's capabilities concretely satisfy the successor's preconditions.

\subsection{RQ4: Comparison with State of the Art}
\label{sec:rq4}

\begin{table*}[t]
\centering
\caption{Performance on component (vulnerability/gadget) discovery and exploit graph construction.}
\renewcommand{\arraystretch}{1.2}
\label{tab:evaluation-main-results}
\begin{adjustbox}{max width=\textwidth}
\setlength{\tabcolsep}{4pt}
\begin{tabular}{ll ccc ccc ccc ccc ccc}
    \toprule
    \multirow{2}{*}{\textbf{Tool}} & \multirow{2}{*}{\textbf{LLM}} & 
    \multicolumn{3}{c}{\textbf{Vulnerability}} &
    \multicolumn{3}{c}{\textbf{Gadget}} &
    \multicolumn{3}{c}{\textbf{Goal Reach Node}} & 
    \multicolumn{3}{c}{\textbf{Edge}} &
    \multicolumn{3}{c}{\textbf{Cond. Goal Reach Edge}} \\
    \cmidrule(lr){3-5} \cmidrule(lr){6-8} \cmidrule(lr){9-11} \cmidrule(lr){12-14} \cmidrule(lr){15-17}
    & & \textbf{TP} & \textbf{FP} & \textbf{Prec. (\%)} & \textbf{TP} & \textbf{FP} & \textbf{Prec. (\%)} & \textbf{TP} & \textbf{FN} & \textbf{Rec. (\%)} & \textbf{TP} & \textbf{FP} & \textbf{Prec. (\%)} & \textbf{TP} & \textbf{FN} & \textbf{Rec. (\%)} \\
    \midrule
    \multirow{3}{*}{\tool} 
        & \textit{DeepSeek-v4-Flash} & 136 & 9  & 93.79\% & 32 & 1  & 96.97\% & 129 & 178 & 42.02\% & 197 & 31  & 86.40\% & 185 & 38 & 82.96\% \\
        & \textit{Qwen-3.8-Flash}     & 132 & 23 & 85.16\% & 46 & 6  & 88.46\% & 132 & 175 & 43.00\% & 273 & 139 & 66.26\% & 107 & 55 & 66.05\% \\
        & \textit{GLM-5.3-Flash}      & 124 & 47 & 72.51\% & 34 & 16 & 68.00\% & 137 & 170 & 44.63\% & 107 & 51  & 67.72\% & 90  & 61 & 59.60\% \\
    \midrule

    \tool$_{\mathrm{PM}}$  & \textit{DeepSeek-v4-Flash} & 89  & 28 & 76.07\% & 8  & 8  & 50.00\% & 79  & 228 & 25.73\% & 18  & 13  & 58.06\% & 12  & 42 & 22.22\% \\
    \tool$_{\mathrm{G}}$   & \textit{DeepSeek-v4-Flash} & 125 & 48 & 72.25\% & 0  & 0  & --       & 99  & 208 & 32.25\% & 167 & 36  & 82.27\% & 68  & 35 & 66.02\% \\
    \tool$_{\mathrm{VK}}$  & \textit{DeepSeek-v4-Flash} & 109 & 20 & 84.50\% & 29 & 2  & 93.55\% & 119 & 188 & 38.76\% & 133 & 11  & 92.36\% & 120 & 36 & 76.92\% \\
    \tool$_{\mathrm{GK}}$  & \textit{DeepSeek-v4-Flash} & 116 & 11 & 91.34\% & 17 & 2  & 89.47\% & 113 & 194 & 36.81\% & 164 & 4   & 97.62\% & 152 & 25 & 85.88\% \\
    \tool$_{\mathrm{TS}}$  & \textit{DeepSeek-v4-Flash} & 104 & 42 & 71.23\% & 9  & 9  & 50.00\% & 86  & 221 & 28.01\% & 25  & 7   & 78.13\% & 13  & 48 & 21.31\% \\
    \midrule
    \multirow{4}{*}{Codex} 
        & \textit{DeepSeek-v4-Flash} & 58  & 9  & 86.57\% & 20 & 11 & 64.52\% & 77  & 230 & 25.08\% & 42  & 2   & 95.45\% & 42  & 28 & 60.00\% \\
        & \textit{Qwen-3.8-Flash}     & 121 & 50 & 70.76\% & 24 & 26 & 48.00\% & 115 & 192 & 37.46\% & 56  & 48  & 53.85\% & 44  & 65 & 40.37\% \\
        & \textit{GLM-5.3-Flash}      & 46  & 11 & 80.70\% & 10 & 3  & 76.92\% & 55  & 252 & 17.92\% & 26  & 9   & 74.29\% & 23  & 15 & 60.53\% \\
        & \textit{GPT-5.4}            & 33  & 5  & 86.84\% & 7  & 6  & 53.85\% & 40  & 267 & 13.03\% & 8   & 6   & 57.14\% & 8   & 10 & 44.44\% \\
    \bottomrule
\end{tabular}
\end{adjustbox}
\end{table*}

\begin{figure}[t]
    \centering
    \includegraphics[width=\columnwidth]{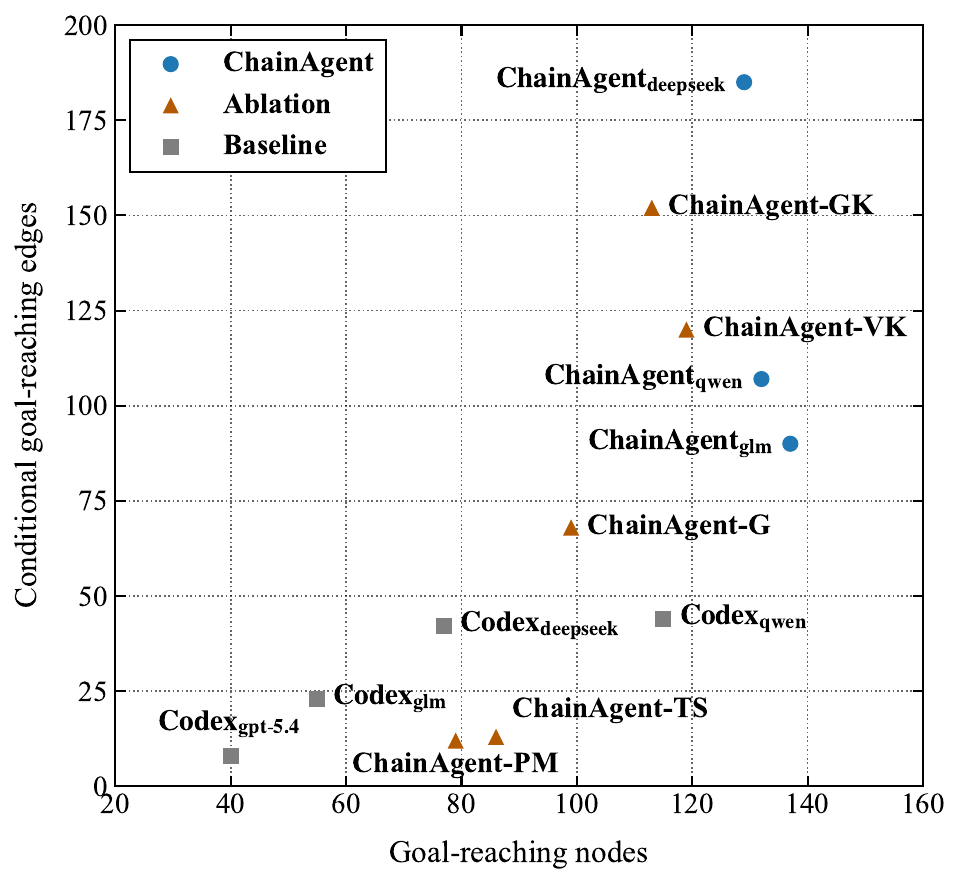}
    \caption{Goal-reaching nodes vs. conditional goal-reaching edges across tool configurations. Points toward the top-right represent superior exploit-graph discovery.}
    \label{fig:coordinate-diagram}
\end{figure}

\noindent\textbf{Overall Performance.}
\Cref{tab:evaluation-main-results} presents the end-to-end performance of \tool and Codex across different LLM backends. Across all three shared models, \tool consistently recovers more goal-reaching nodes and valid edges than Codex. Its advantage is particularly pronounced in edge recovery: \tool constructs $2.4$--$4.4\times$ as many correct conditional goal-reaching edges under the same backend, covering substantially more of the verified capability transfers toward the security goal.

For component discovery, \tool identifies 136 true-positive vulnerabilities and 32 functional gadgets under \textit{DeepSeek-v4-Flash}, compared with Codex's 58 vulnerabilities and 20 gadgets. This corresponds to $2.3\times$ as many vulnerabilities and $1.6\times$ as many gadgets, with only 9 and 1 false positives, respectively, yielding precision of 93.79\% and 96.97\%. Under \textit{Qwen-3.8-Flash}, the difference in vulnerability counts is smaller (132 vs.\ 121), but \tool discovers nearly twice as many functional gadgets (46 vs.\ 24). For both backends, \tool achieves higher precision for both component types alongside these gains in discovery. Under \textit{GLM-5.3-Flash}, \tool also finds substantially more true-positive vulnerabilities (124 vs.\ 46) and gadgets (34 vs.\ 10), although Codex achieves higher precision for both types. Thus, the increase in valid component discoveries holds across all three models, while the precision advantage depends on the backend.

The performance gap widens dramatically in exploit graph construction. For goal-reaching components, \tool achieves a stable recall of 42.02\%--44.63\%, consistently outpacing Codex (13.03\%--37.46\%). For edge construction, \tool achieves precision of 86.40\%, 66.26\%, and 67.72\% under \textit{DeepSeek}, \textit{Qwen}, and \textit{GLM}, respectively, compared with Codex's 95.45\%, 53.85\%, and 74.29\%. Crucially, on Cond. Goal Reach Edge, which evaluates the ability to connect discovered components into valid attack paths, \tool achieves up to 82.96\% recall (recovering 185 ground-truth edges under \textit{DeepSeek-v4-Flash}), compared to Codex's 60.00\% recall (recovering only 42 edges). Notably, even when Codex is powered by its officially recommended model, \textit{GPT-5.4}, its discovery performance remains severely constrained (achieving only 13.03\% Goal Reach recall and recovering 8 conditional edges). These results demonstrate that general-purpose coding agents struggle to autonomously reason about exploit dependencies, whereas \tool's goal-directed architecture enables systematic and effective exploit graph construction.

To illustrate the graph-recovery advantage, \Cref{fig:coordinate-diagram} plots discovered goal-reaching nodes against conditional goal-reaching edges across all configurations. As shown, most \tool configurations dominate the baseline: while Codex can occasionally discover standalone nodes (e.g., 115 nodes with \textit{Qwen}), it remains trapped near the horizontal axis with fewer than 45 edges. In contrast, \tool successfully translates discovered components into interconnected exploit paths, with $\mathrm{ChainAgent}_{\mathrm{deepseek}}$ establishing the top-right performance frontier.

\begin{figure*}[t]
    \centering
    \includegraphics[width=0.95\textwidth]{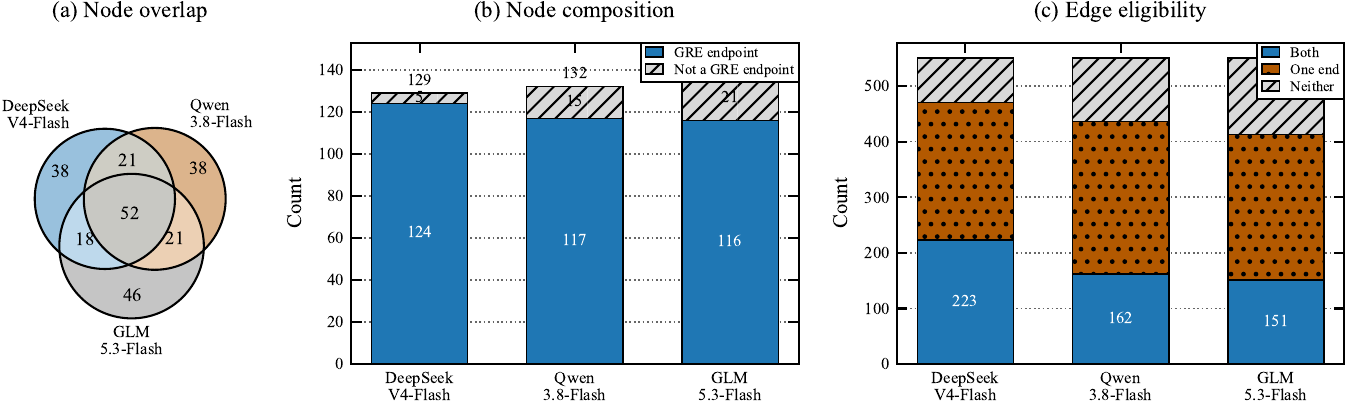}
    \caption{In-depth comparison of \tool across three LLM backends: (a) overlap of discovered goal-reaching nodes, (b) breakdown of goal-reaching nodes into GRE endpoints and non-endpoints, and (c) edge eligibility based on whether both, one, or neither endpoint is discovered.}
    \label{fig:goal-reach-panels}
\end{figure*}

\noindent\textbf{Impact of LLM Backends.}
Although \tool achieves similar Goal Reach node counts across the three backends (129--137 in \Cref{tab:evaluation-main-results}), their edge construction performance diverges sharply. \Cref{fig:goal-reach-panels} inspects the node distribution and structural cohesiveness underlying this divergence. \ding{182} \textbf{Model Complementarity}. The Venn diagram in \Cref{fig:goal-reach-panels} (a) demonstrates strong complementarity across models: only 52 goal-reaching nodes are shared by all three backends, while each uniquely discovers 38 (\textit{DeepSeek}), 38 (\textit{Qwen}), and 46 (\textit{GLM}) nodes. In total, their union covers 234 goal-reaching nodes, indicating that different LLMs explore markedly distinct attack paths. \ding{183} \textbf{Structural Cohesiveness and Edge Eligibility}. \Cref{fig:goal-reach-panels} (b) and (c) reveal why similar node totals yield disparate edge recall. \Cref{fig:goal-reach-panels} (b) shows that 124 of \textit{DeepSeek}'s 129 nodes (96.1\%) are valid Goal Reaching Edge (GRE) endpoints, with only 5 dangling nodes, whereas \textit{Qwen} and \textit{GLM} produce 15 and 21 dangling nodes, respectively. Consequently, as shown in \Cref{fig:goal-reach-panels} (c), \textit{DeepSeek}'s tightly coupled discoveries unlock \textit{both endpoints} for 223 ground-truth edges, substantially surpassing \textit{Qwen} (162) and \textit{GLM} (151), where most findings cover only a single endpoint. By prioritizing interconnected components over isolated flaws, \textit{DeepSeek-V4-Flash} establishes a vastly superior foundation for synthesizing valid exploit graphs.

\begin{table}[t]
\centering
\caption{Token consumption (in millions) and execution time across benchmark applications.}
\label{tab:evaluation-cost}
\begin{adjustbox}{width=\columnwidth}
\begin{tabular}{ll rrrr}
    \toprule
    \multirow{2}{*}{\textbf{Tool}} & \multirow{2}{*}{\textbf{LLM}} & \multicolumn{3}{c}{\textbf{Tokens (M)}} & \multirow{2}{*}{\textbf{Time}} \\
    \cmidrule(lr){3-5}
    & & \textbf{In} & \textbf{Out} & \textbf{Total} & \\
    \midrule
    \multirow{3}{*}{\tool} 
        & \textit{DeepSeek-v4-Flash} & 442.80  & 3.91  & 446.71  & 7h28m44s  \\
        & \textit{Qwen-3.8-Flash}     & 1399.10 & 17.86 & 1416.96 & 35h52m44s \\
        & \textit{GLM-5.3-Flash}      & 885.58  & 2.68  & 888.26  & 28h43m29s \\
    \midrule
    \multirow{4}{*}{Codex} 
        & \textit{Deepseek-v4-Flash} & 216.88  & 0.98  & 217.86  & 3h1m4s    \\
        & \textit{Qwen-3.8-Flash}     & 170.78  & 1.31  & 172.09  & 9h0m52s   \\
        & \textit{GLM-5.3-Flash}      & 109.40  & 0.50  & 109.90  & 8h5m46s   \\
        & \textit{GPT-5.4}            & 114.08  & 1.28  & 115.36  & 8h44m42s  \\
    \bottomrule
\end{tabular}
\end{adjustbox}
\end{table}

\noindent\textbf{Cost Analysis.}
\Cref{tab:evaluation-cost} reports the total token consumption and execution time across all 15 benchmark applications. 
While \tool incurs higher overhead than Codex, this cost directly fuels its iterative, goal-directed exploration; in contrast, Codex terminates prematurely without systematically pursuing multi-step paths. 
Crucially, under \textit{deepseek-v4-flash}, \tool achieves its peak exploit-graph construction performance at a highly practical cost: 446.71M tokens and 7.5 hours in total, averaging under 30 minutes per real-world application. 
Meanwhile, \textit{qwen-3.8-flash} performs the most exhaustive reasoning (consuming 17.86M output tokens over 35.9 hours), which accounts for its superior gadget discovery but requires a higher compute budget.

\begin{rqbox}
\noindent\textbf{Answer to RQ4.} 
\tool substantially outperforms Codex, discovering up to $2.3\times$ more vulnerabilities and recovering $4.4\times$ as many correct conditional goal-reaching edges (185 vs. 42), effectively constructing goal-directed exploit graphs at practical cost.
\end{rqbox}

\subsection{RQ5: Ablation Studies}
\label{sec:rq5}

\noindent\textbf{Ablation Setup.} To isolate the contribution of each core component in \tool, we evaluate five ablation variants under \textit{DeepSeek-v4-Flash}:
\begin{itemize}
    \item \textbf{\tool$_{\mathrm{PM}}$:} Disables explicit precondition--capability modeling, relying on generic LLM prompts to hypothesize connections.
    \item \textbf{\tool$_{\mathrm{G}}$:} Excludes functional gadgets, restricting exploration strictly to unintended vulnerabilities.
    \item \textbf{\tool$_{\mathrm{VK}}$:} Removes the empirical knowledge guiding vulnerability searches from preconditions.
    \item \textbf{\tool$_{\mathrm{GK}}$:} Removes the empirical knowledge guiding gadget searches.
    \item \textbf{\tool$_{\mathrm{TS}}$:} First independently searches for vulnerabilities throughout the application using the same auditing mechanisms as \textsc{FindVuln} agent, retaining valid flaws even if their limited capabilities cannot achieve RCE alone. It then fixes the vulnerability set, infers preconditions and capabilities, and reuses gadget search, precondition--capability matching, and chain synthesis, without further \textsc{FindVuln} calls for unmet preconditions.
\end{itemize}

We evaluate these configurations across the same metrics, presenting quantitative results in \Cref{tab:evaluation-main-results} and their divergence from the optimal frontier in \Cref{fig:coordinate-diagram}.

\noindent\textbf{Impact of Precondition Model and Functional Gadgets.}
The results confirm that the precondition model is the primary engine of graph construction, while functional gadgets provide essential transitions for attack paths. 
Specifically, removing the precondition model (\tool$_{\mathrm{PM}}$) causes the most severe performance collapse: goal-reaching node recall drops from 42.02\% to 25.73\% (TP from 129 to 79), edge precision falls from 86.40\% to 58.06\%, and conditional edge recall plummets from 82.96\% to a meager 22.22\% (with only 12 edges recovered). As visible in Figure~\ref{fig:coordinate-diagram}, \tool$_{\mathrm{PM}}$ falls to the very bottom of the plot. Without structured capability matching, the agent cannot track capability dependencies, rendering it unable to verify whether adjacent components compose.
Similarly, omitting functional gadgets (\tool$_{\mathrm{G}}$) sharply curtails the attack surface, reducing goal-reaching nodes to 99 and edges from 185 to 68. Because individual web vulnerabilities frequently grant only non-executable intermediate states, functional gadgets serve as indispensable bridges to translate these states into critical security goals.

\noindent\textbf{Impact of Chaining Search Knowledge.}
Our empirical chaining knowledge is critical for focusing LLM reasoning and maximizing search recall. 
Removing vulnerability search knowledge (\tool$_{\mathrm{VK}}$) causes vulnerability TP to drop from 136 to 109 while more than doubling false positives (from 9 to 20), lowering recovered edges to 120. Without guidance mapping preconditions to likely sinks, the agent wastes reasoning budget on irrelevant code paths, leading to missed flaws and hallucinated bugs. 
Similarly, removing gadget search knowledge (\tool$_{\mathrm{GK}}$) nearly halves discovered gadgets from 32 to 17 (dropping goal-reaching nodes to 113). Since intended program behaviors lack overt vulnerability sinks, unguided search struggles across the vast codebase. As shown in Figure~\ref{fig:coordinate-diagram}, both knowledge ablations shift the agent downward and leftward, demonstrating that domain knowledge is vital for comprehensive component discovery.

\noindent\textbf{Joint Discovery vs.\ Discovery-then-Plan.}
\tool$_{\mathrm{TS}}$ separates vulnerability discovery from chain construction while retaining the same precondition--capability model, gadget search, and composition mechanisms. Joint discovery correctly identifies 108 vulnerabilities with a path to the goal in the ground-truth graph, compared with 79 for \tool$_{\mathrm{TS}}$: they share 50, with 58 unique to \tool and 29 unique to \tool$_{\mathrm{TS}}$. With a fixed vulnerability set, \tool$_{\mathrm{TS}}$ cannot use unmet preconditions exposed during planning to resume vulnerability search. \tool instead turns these dependencies into focused source-code search tasks, allowing newly discovered components to guide further exploration. Under \tool$_{\mathrm{TS}}$, goal-reaching node recall falls from 42.02\% to 28.01\%, and conditional edge recall drops from 82.96\% to 21.31\%. Edge precision also decreases from 86.40\% to 78.13\%. Together, these results support our motivation that separating vulnerability discovery from planning can miss components that contribute to exploit chains.

\begin{rqbox}
\noindent\textbf{Answer to RQ5.} 
The precondition model and functional gadgets are indispensable for establishing exploit-chain edges, while empirical chaining knowledge effectively focuses the search space and maximizes component discovery recall. Joint discovery further improves graph recovery by using emerging dependencies to guide continued component search.
\end{rqbox}

\subsection{RQ6: Real-World Applications}
\label{sec:rq6}

\noindent\textbf{Application Selection.} We select 30 real-world PHP Web applications from GitHub repositories with more than 1,000 stars, excluding SDKs, command-line tools, libraries, and other non-web projects. For each subject, we analyze the latest available version at the time of collection.

\noindent\textbf{Findings.} As shown in Table~\ref{tab:real-world-findings}, \tool discovers 20 vulnerabilities in seven of the 30 applications. These vulnerabilities form a total of 13 exploit chains, 9 of which achieve RCE. Functional gadgets are involved in 9 of these chains. All identified vulnerabilities and exploit chains have been dynamically validated and reported to the affected developers, with 9 vulnerabilities receiving CVE-IDs.

\noindent\textbf{Case Study 1: Configuration Overwrite to RCE in Open Web Analytics (OWA).} OWA is a web-analytics application. An authenticated attacker with the \texttt{edit\_settings} capability and a valid options-form nonce can turn a configuration overwrite into RCE through normal error logging. Appendix~\Cref{lst:case-owa} shows the relevant code.

The options handler checks permission to edit settings but does not restrict which setting names the request may persist. The attacker can therefore overwrite both \texttt{report\_wrapper} and \texttt{error\_log\_file}. The wrapper name appears in the error message when a template cannot be found, while the log setting determines where that message is written. The attacker stores a quote-free, quoted-printable-encoded PHP fragment as the nonexistent wrapper name and sets the log destination to a \texttt{php://filter} stream that applies \texttt{convert.quoted-printable-decode} before writing to a PHP file at an attacker-chosen path. Encoding the angle brackets lets the fragment survive input cleaning. A subsequent page render triggers the missing-template error, and the logger writes the message through the decoding filter. If the destination is writable and web-accessible with PHP execution enabled, requesting the resulting file achieves RCE. The logging gadget thus turns attacker-controlled settings into executable file contents, completing $V_{\mathit{config}} \rightarrow G_{\mathit{logging}} \rightarrow \mathrm{RCE}$.

\noindent\textbf{Case Study 2: Cross-User Credential Theft in TeamPass.} TeamPass is a collaborative password manager. Its import workflow first stores an uploaded file, then parses its contents into staged rows, and finally creates password items. An ordinary authenticated import user can cross these request boundaries to convert a victim's pending CSV upload into attacker-readable credentials. The attack assumes that import is enabled, the victim has not yet consumed the upload, and the attacker obtains the numeric upload \texttt{operation\_id}. Appendix~\Cref{lst:case-teampass} gives a simplified view of the vulnerable workflow.

The upload handler records the temporary filename and returns a global auto-increment identifier without binding that record to its uploader. The CSV parser later retrieves the file using only this identifier and its \texttt{temp\_file} type. By supplying a victim's pending identifier, the attacker makes the parser read the victim's file and stage its password rows under the same \texttt{operation\_id}. The attacker then submits that identifier as \texttt{csvOperationId} and selects their own personal folder with \texttt{folderId}. For CSV rows without an assigned folder, the normal import workflow encrypts the staged passwords, creates items in the selected folder, and shares their keys with the caller. The import gadget therefore converts the victim's uploaded data into password items the attacker can read, completing $V_{\mathit{takeover}} \rightarrow G_{\mathit{import}} \rightarrow \mathrm{credential\ theft}$.

\noindent\textbf{Case Study 3: SQL Injection to RCE in Ecommerce CodeIgniter Bootstrap.} Ecommerce CodeIgniter Bootstrap is an e-commerce application. This chain requires an administrator session, an order-processing event for an attacker-controlled order, and writable PHP error logging. The order's stored quantity reaches a SQL query, failed queries are logged verbatim, and the language editor includes PHP files from a registered language directory. Appendix~\Cref{lst:case-ecommerce} shows the relevant code.

An attacker can submit an order whose quantity contains PHP code; when an administrator processes it, the value is concatenated into the product-update query. The query fails, and CodeIgniter's database driver writes the invalid SQL, including the injected bytes, to its PHP error log. With an administrator session, the attacker first registers a language named \texttt{../logs}, then triggers the failed query: language registration creates the target directory, so this ordering preserves the subsequently poisoned log. Opening that registered language makes the editor traverse into the log directory and require its PHP files. Because the include runs inside the bootstrapped application, the log's \texttt{BASEPATH} guard is already satisfied; the injected PHP code executes during inclusion. The SQL injection places code in the log, the intended logging function carries it to a PHP file, and the file-inclusion vulnerability executes it, completing $V_{\mathit{SQLi}} \rightarrow G_{\mathit{logging}} \rightarrow V_{\mathit{include}} \rightarrow \mathrm{RCE}$.

\begin{rqbox}
\noindent\textbf{Answer to RQ6.} \tool discovers 20 vulnerabilities across seven real-world applications and identifies 13 exploit chains, nine of which achieve RCE.
\end{rqbox}

\section{Discussion}
\label{sec:discussion}

\noindent\textbf{Threats to Validity.} Our ground truth combines findings from all evaluated tools with documented vulnerabilities. Chains missed by every tool may remain absent, so recall is measured against the collected and validated set rather than all exploitable chains in an application.

\noindent\textbf{Limitations.} Although we validate findings separately during evaluation, \tool does not generate and execute exploits within its discovery pipeline. During early development, we explored integrating dynamic validation into the discovery loop using an agent based on \textit{GPT-5.4} to automatically deploy the target application in Docker and generate and test an exploit. From environment setup to a working exploit, this process took over an hour on average, substantially slowing the discovery loop. Repeating such checks for every candidate vulnerability, gadget, and edge would make exploration increasingly expensive as the graph grows. Interleaving exploit generation and debugging with discovery can also crowd the agent's context with payload attempts and execution logs, leaving less room for code evidence and capability dependencies. We therefore keep validation outside the discovery loop to preserve scalability, but the pipeline itself does not confirm whether a complete chain executes successfully.

\noindent\textbf{Future Work.} We plan to investigate graph-guided dynamic validation that first checks nodes shared by many goal-reaching paths or connecting large subgraphs. If validation rules out a required capability under the stated assumptions, we can remove the edges that depend on it. This may split the graph into smaller subgraphs, allowing us to skip regions that no longer lie on any remaining entry-to-goal path. An unsuccessful exploit attempt alone would not justify pruning. We will also explore reusing environment snapshots and validation results for shared chain prefixes, with a separate validation context returning concise evidence to the discovery process.

\section{Related Work}
\label{related-work}

\noindent\textbf{Automated Attack Planning and Penetration Testing.} Traditional penetration testing relies heavily on human expertise to manually uncover and chain vulnerabilities from a black-box perspective, which is effective yet labor-intensive and hard to scale. To automate attack planning and execution, recent research explores LLM-based autonomous agents. Early efforts such as PentestGPT~\cite{DBLP:conf/uss/PentestGPT} mitigate context loss via modular reasoning, while subsequent systems enhance autonomy across reconnaissance and exploitation via specialized multi-agent collaboration~\cite{DBLP:conf/asiaccs/PentestAgent}. Other systems use task dependency graphs and structured memory to manage longer engagements~\cite{DBLP:conf/ijcai/PENTESTLLMAGENT, DBLP:journals/eswa/PenExpert}. For multi-host red teaming, Incalmo~\cite{DBLP:conf/sp/Incalmo} delegates the execution of high-level plans to domain-specific agents.

Another line of automated attack planning uses explicit action models to compose known exploits and system operations toward security objectives.
Foundational work formalized network penetration testing as planning problems with explicit action requirements and state transitions~\cite{DBLP:journals/corr/ObesSR13}. Modern systems specialize this concept for specific attack phases: ChainReactor~\cite{DBLP:conf/uss/ChainReactor} formulates single-host post-exploitation privilege escalation (PE) as a PDDL planning problem, synthesizing paths from a low-privileged shell to root by combining known CVEs with benign system operations.
At the network level, ALFA-Chains~\cite{DBLP:journals/corr/ALFA-Chains} plans multi-host hybrid attack paths across known topologies, scheduling known PE and RCE exploits to facilitate lateral movement toward a target host.

However, these planning frameworks operate post-hoc over a closed world of known vulnerabilities, focusing narrowly on single-host PE or cross-host PE/RCE lateral movement without discovering new flaws. In contrast, \tool integrates compositional exploit reasoning directly into source-code-level vulnerability discovery for web applications. Guided by unmet preconditions and generated capabilities in an exploit graph, \tool actively uncovers previously unknown vulnerabilities and functional gadgets across arbitrary flaw types, chaining them to achieve critical security goals.

\noindent\textbf{Exploit Generation for Single Vulnerabilities.} One line of work generates exploits for taint-style injection vulnerabilities. For PHP applications, ChainSaw~\cite{DBLP:conf/ccs/Chainsaw} combines symbolic execution with workflow analysis to generate inputs reaching vulnerable sinks, while NAVEX~\cite{alhuzali2018navex} integrates dynamic crawling to handle client-side execution. In Node.js, FAST~\cite{DBLP:conf/sp/FAST} uses taint-guided constraint solving, and Explode.js~\cite{DBLP:journals/pacmpl/Explode.js} constructs call sequences reaching vulnerable sinks. NodeMedic~\cite{DBLP:conf/eurosp/NodeMedic} and NodeMedic-FINE~\cite{DBLP:conf/ndss/NodeMedic-FINE} combine dynamic taint tracking with fuzzing to generate structured payloads for code and command injection.

Other approaches target unsafe deserialization and regular expression denial of service (ReDoS). FUGIO~\cite{DBLP:conf/uss/FUGIO} combines gadget-chain analysis with fuzzing to generate PHP exploit objects, while PFortifier~\cite{DBLP:conf/sp/PFortifier} analyzes object-injection gadget chains to guide patching. ReDoSHunter~\cite{DBLP:conf/uss/ReDoSHunter} and RENGAR~\cite{DBLP:conf/sp/RENGAR} model vulnerable regex patterns and generate strings that trigger super-linear matching costs; EvilStrGen~\cite{DBLP:conf/uss/EVILSTRGEN} addresses non-backtracking engines, and PufferDoS~\cite{DBLP:conf/sp/XuDCLSTKM26} refines and validates strings against program-level constraints.

Complementing these analyses, PoCGen~\cite{DBLP:journals/pacmse/PoCGen} combines LLM reasoning with static taint tracking and runtime feedback to generate executable PoCs for npm vulnerabilities described in CVE reports.

Exploit generation in these settings focuses on triggering individual vulnerabilities, even when exploitation traverses multiple calls or deserialization gadgets. In contrast, \tool uses a precondition--capability model to jointly discover vulnerabilities and functional gadgets and construct dependencies, revealing how limited capabilities combine across components to reach a security goal.

\section{Conclusion}
\label{sec:conclusion}

We presented \tool, a goal-directed LLM-agent framework for exploit-chain discovery in PHP applications. Grounded in our empirical findings, \tool models vulnerabilities and functional gadgets through preconditions and capabilities, and uses object-centered chaining knowledge to guide discovery and composition. On 15 benchmark applications, \tool recovers up to $4.4\times$ as many correct conditional goal-reaching edges as Codex under the same LLM backend. An additional evaluation of 30 real-world applications documents 20 vulnerabilities and 13 exploit chains, nine of which achieve RCE.

%
\bibliographystyle{IEEEtran}
\bibliography{ref}

\newpage
\appendix
\label{sec:appendix}
\renewcommand{\thesubsectiondis}{\thesubsection.}
\subsection{Ethics Considerations}
\label{sec:ethics}

\noindent\textbf{Responsible Disclosure.} We contacted affected maintainers through their repositories' designated security-reporting channels, including email and private GitHub security advisories, and coordinated with VulDB and the official CVE request service. Our reports included vulnerability details and recommended fixes. For newly reported findings, we withheld our own public disclosures for at least 90 days after initial notification to give maintainers time to prepare patches. At the time of writing, nine of the reported vulnerabilities have been assigned CVE identifiers.

\noindent\textbf{Dual-Use Risks.} \tool could lower the effort required to find attack paths leading to RCE or data theft. Its pipeline statically analyzes source code to identify exploit chains along with their entry assumptions. To mitigate potential misuse, we do not publicly release the source code of \tool. We further follow responsible disclosure practices and the publication delays described above to give affected projects sufficient time to address the reported weaknesses.

\subsection{CWE Co-occurrence}

\begin{figure}[!ht]
    \centering
    \includegraphics[width=\columnwidth]{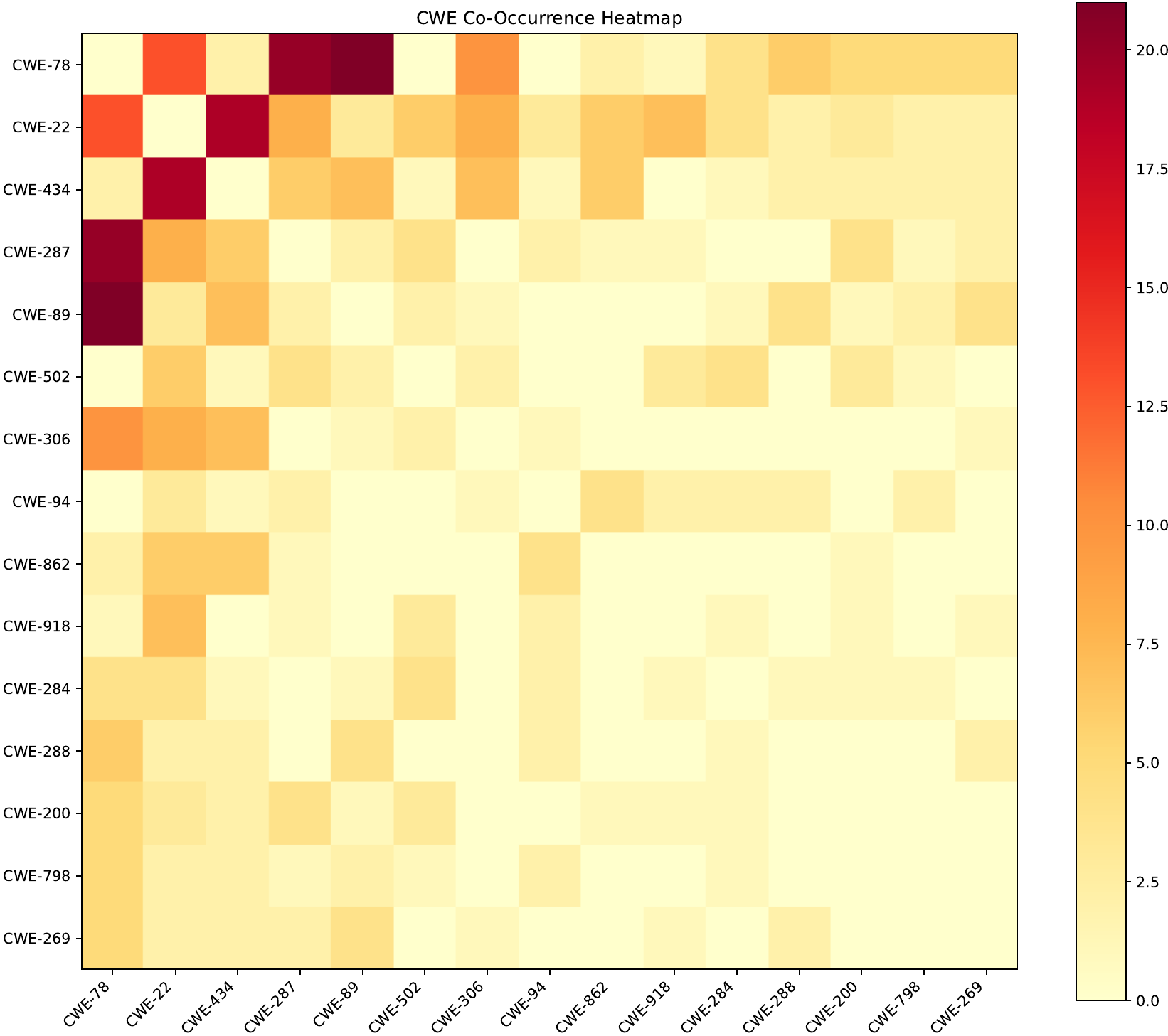}
    \caption{Co-occurrence heatmap of the top-15 most frequent CWEs in the collected exploit chains.}
    \label{fig:vuln-heatmap}
\end{figure}

\newpage

\subsection{Real-World Vulnerability Findings}

\noindent
\begin{minipage}{\columnwidth}
\begin{lstlisting}[style=phpStyle, label={lst:case-owa}, caption={OWA: request-controlled settings flow into the template-error message and the logger's output stream.}]
// (1) optionsUpdate.php: settings update
$this->setRequiredCapability('edit_settings');
// ...
$c = owa_coreAPI::configSingleton();
$config_values = $this->get('config');
// Inside the nonempty-config branch:
foreach ($config_values as $k => $v) {
    list($module, $name) = explode('.', $k);
    if ($module && $name) {
        $c->persistSetting($module, $name, $v);
    }
}
$c->save();

// (2) owa_view.php: select the outer template
$this->t->set_template(
    $this->config['report_wrapper']);
// owa_template.php: template-not-found branch
$this->e->err(sprintf(
    '%s was not found in any template directory.',
    $file));

// (3) classes/error.php: configure the log sink
$path = owa_coreAPI::getSetting(
    'base', 'error_log_file');
$stream = new StreamHandler($path, $level);
// ... formatter setup omitted ...
$this->logger->pushHandler($stream);
\end{lstlisting}
\end{minipage}
\par\medskip

\noindent
\begin{minipage}{\columnwidth}
\begin{lstlisting}[style=phpStyle, label={lst:case-teampass}, caption={TeamPass: simplified upload lookup and CSV-to-item conversion.}]
// (1) Upload lookup: no owner check
$op = $inputData['file'];
$upload = DB::queryFirstRow(
    'SELECT valeur FROM misc
     WHERE increment_id=%i
       AND type="temp_file"', $op);
// Parse CSV; stage each password:
DB::insert('items_importations', [
    'operation_id' => $op, 'pwd' => $pwd
]);

// (2) Import rows using the same ID
$rows = DB::query(
    'SELECT pwd FROM items_importations
     WHERE operation_id=%i',
    $dataReceived['csvOperationId']);
foreach ($rows as $row) {
    $enc = doDataEncryption($row['pwd']);
    DB::insert('items', [
        'pw' => $enc['encrypted'],
        // CSV row has no assigned folder:
        'id_tree' => $dataReceived['folderId']
    ]);
    $id = DB::insertId();
    // Personal folder: grant caller the key
    storeUsersShareKey(
        'sharekeys_items', 1, $id,
        $enc['objectKey'], true);
}
\end{lstlisting}
\end{minipage}
\par\medskip

\begin{table*}[!t]
    \centering
    \caption{RQ6: vulnerabilities and exploit chains in real-world PHP applications.}
    \label{tab:real-world-findings}
    \footnotesize
    \setlength{\tabcolsep}{3.5pt}
    \renewcommand{\arraystretch}{1.1}
    \begin{tabularx}{\textwidth}{@{}>{\raggedright\arraybackslash}p{0.17\textwidth}rcc>{\raggedright\arraybackslash}Xl>{\raggedright\arraybackslash}p{0.34\textwidth}@{}}
        \toprule
        \multirow{2}{*}{\textbf{Application}} &
        \multirow{2}{*}{\textbf{\#Stars}} &
        \multicolumn{4}{c}{\textbf{Vulnerabilities}} &
        \multirow{2}{*}{\textbf{Exploit Chains}} \\
        \cmidrule(lr){3-6}
        & & \textbf{No.} & \textbf{ID} & \textbf{Type} & \textbf{Status} & \\
        \midrule

        \multirow{3}{=}{\raggedright\bfseries Ecommerce CodeIgniter Bootstrap}
        & \multirow{3}{*}{1,437}
        & \#1 & $V_1$ & Code injection & \multirow{3}{*}{CVE-ID Assigned} & \multirow{3}{*}{%
            \shortstack[l]{%
                $V_2 \rightarrow G_{\mathit{logging}} \rightarrow V_3 \rightarrow \mathrm{RCE}$
            }} \\
        & & \#2 & $V_2$ & SQLi & & \\
        & & \#3 & $V_3$ & File inclusion & & \\

        \midrule

        \multirow{4}{=}{\raggedright\bfseries GEOFlow}
        & \multirow{4}{*}{4,038}
        & \#4 & $V_1$ & XSS & \multirow{4}{*}{CVE-ID Assigned} &
        \multirow{4}{*}{%
            \shortstack[l]{%
                $V_1 \rightarrow V_2 \rightarrow \mathrm{file\ deletion}$\\
                $V_1 \rightarrow V_3 \rightarrow \mathrm{RCE}$\\
                $V_1 \rightarrow V_4 \rightarrow \mathrm{SSRF}$
            }} \\
        & & \#5 & $V_2$ & File deletion & & \\
        & & \#6 & $V_3$ & Blade SSTI & & \\
        & & \#7 & $V_4$ & SSRF & & \\

        \midrule

        \multirow{2}{=}{\raggedright\bfseries GitList}
        & \multirow{2}{*}{2,948}
        & \#8 & $V_1$ & Git trust bypass & \multirow{2}{*}{CVE-ID Assigned} &
        \multirow{2}{*}{%
            \shortstack[l]{%
                $V_1 \rightarrow G_{\mathit{textconv}} \rightarrow \mathrm{RCE}$\\
                $V_2 \rightarrow G_{\mathit{XML}} \rightarrow \mathrm{persistent\ DoS}$
            }} \\
        & & \#9 & $V_2$ & XML injection & & \\

        \midrule

        \multirow{3}{=}{\raggedright\bfseries MacCMS}
        & \multirow{3}{*}{2,924}
        & \#10 & $V_1$ & Privilege escalation & \multirow{3}{*}{Fixed} &
        \multirow{3}{*}{%
            \shortstack[l]{%
                $V_1 \rightarrow V_2 \rightarrow G_{\mathit{mail}} \rightarrow \mathrm{RCE}$\\
                $V_1 \rightarrow V_3 \rightarrow G_{\mathit{static}} \rightarrow \mathrm{RCE}$
            }} \\
        & & \#11 & $V_2$ & Stored SSTI & & \\
        & & \#12 & $V_3$ & Template traversal & & \\

        \midrule

        \textbf{Open Web Analytics}
        & 2,697
        & \#13 & $V_1$ & Config overwrite & Fixed &
        $V_1 \rightarrow G_{\mathit{logging}} \rightarrow \mathrm{RCE}$ \\

        \midrule

        \multirow{5}{=}{\raggedright\bfseries Paymenter}
        & \multirow{5}{*}{2,831}
        & \#14 & $V_1$ & API-scope escalation & \multirow{5}{*}{Fixed} &
        \multirow{5}{*}{%
            \shortstack[l]{%
                $V_1 \rightarrow V_2 \rightarrow V_3 \rightarrow \mathrm{RCE}$\\
                $V_1 \rightarrow V_2 \rightarrow V_4 \rightarrow G_{\mathit{mail}} \rightarrow \mathrm{RCE}$\\
                $V_1 \rightarrow V_2 \rightarrow V_5 \rightarrow G_{\mathit{notify}} \rightarrow \mathrm{RCE}$
            }} \\
        & & \#15 & $V_2$ & Role escalation & & \\
        & & \#16 & $V_3$ & Extension RCE & & \\
        & & \#17 & $V_4$ & Mail SSTI & & \\
        & & \#18 & $V_5$ & Product SSTI & & \\

        \midrule

        \multirow{2}{=}{\raggedright\bfseries TeamPass}
        & \multirow{2}{*}{1,890}
        & \#19 & $V_1$ & Import takeover & \multirow{2}{*}{Fixed} &
        \multirow{2}{*}{%
            $\displaystyle V_1 \rightarrow G_{\mathit{import}} \rightarrow \mathrm{credential\ theft}$} \\
        & & \#20 & $V_2$ & Folder ACL bypass & & \\

        \midrule

        \textbf{Total}
        & -- & \multicolumn{4}{l}{\textbf{20 vulns / 7 apps}}
        & \textbf{13 chains / 7 apps} \\

        \bottomrule
    \end{tabularx}
\end{table*}

\noindent
\begin{minipage}{\columnwidth}
\begin{lstlisting}[style=phpStyle, label={lst:case-ecommerce}, caption={Ecommerce CodeIgniter Bootstrap: SQL-to-log-to-include chain.}]
// (1) Process an order with a stored quantity
$products = unserialize($order['products']);
$product = $products[0];
$sql = 'UPDATE products SET quantity=quantity-'
    . $product['product_quantity']
    . ' WHERE id='
    . $product['product_info']['id'];
$this->db->query($sql);

// (2) DB_driver.php logs failed SQL verbatim
log_message('error', 'Query error: '
    . $error['message']
    . ' - Invalid query: ' . $sql);
// application/logs/log-YYYY-MM-DD.php

// (3) Languages.php includes registered PHP files
$name = $_GET['editLang'];
if ($this->Languages_model->countLangs(
        $name) == 0) {
    redirect('admin/languages');
}
$path = 'application/language/' . $name . '/';
$files = rreadDir($path);
foreach ($files['php'] as $file) {
    require $file;
}
\end{lstlisting}
\end{minipage}
\par\medskip

\end{document}